\documentclass[12pt,preprint]{emulateapj}
\usepackage{graphicx}

\shorttitle{Orbital Dynamics of Multi-Planet Systems}
\shortauthors{Stephen R. Kane \& Sean N. Raymond}
\slugcomment{Submitted for publication in the Astrophysical Journal}

\begin{document}

\title{Orbital Dynamics of Multi-Planet Systems with Eccentricity
  Diversity}
\author{
  Stephen R. Kane\altaffilmark{1},
  Sean N. Raymond\altaffilmark{2,3}
}
\email{skane@sfsu.edu}
\altaffiltext{1}{Department of Physics \& Astronomy, San Francisco
  State University, 1600 Holloway Avenue, San Francisco, CA 94132,
  USA}
\altaffiltext{2}{CNRS, UMR 5804, Laboratoire d'Astrophysique de
  Bordeaux, 2 rue de l'Observatoire, BP 89, F-33271 Floirac Cedex,
  France}
\altaffiltext{3}{Universit\'e de Bordeaux, Observatoire Aquitain des
  Sciences de l'Univers, 2 rue de l'Observatoire, BP 89, F-33271
  Floirac Cedex, France}


\begin{abstract}

Since exoplanets were detected using the radial velocity method, they
have revealed a diverse distribution of orbital
configurations. Amongst these are planets in highly eccentric orbits
($e > 0.5$). Most of these systems consist of a single planet but
several have been found to also contain a longer period planet in a
near-circular orbit. Here we use the latest Keplerian orbital
solutions to investigate four known systems which exhibit this extreme
eccentricity diversity; HD~37605, HD~74156, HD~163607, and
HD~168443. We place limits on the presence of additional planets in
these systems based on the radial velocity residuals. We show that the
two known planets in each system exchange angular momentum through
secular oscillations of their eccentricities. We calculate the
amplitude and timescale for these eccentricity oscillations and
associated periastron precession. We further demonstrate the effect of
mutual orbital inclinations on the amplitude of high-frequency
eccentricity oscillations. Finally, we discuss the implications of
these oscillations in the context of possible origin scenarios for
unequal eccentricities.

\end{abstract}



\section{Introduction}
\label{intro}

The discovery of exoplanets has yielded many surprises with regards to
their properties in comparison with the planets in our Solar
System. Amongst these are planets in highly eccentric orbits ($e >
0.5$), such as 16~Cyg~B~b \citep{coc97} and HD~80606b \citep{nae01}.
The number of detected exoplanets is sufficient to note a divergence
from circular orbits which occurs beyond a semi-major axis of $\sim
0.1$~AU \citep{but06}. There is evidence that this correlation is also
present in the exoplanetary candidates discovered by the Kepler
mission \citep{kan12a}, along with evidence that eccentricities tend
to decrease with decreasing planetary size. For planets below the size
of Neptune, orbits are more likely to be circular since these planets
cannot efficiently excite the eccentricities of other planets so
collisions are favored over scattering \citep{gol04}. The more
efficient tidal dissipation in this regime also leads to shorter tidal
circularization time scales \citep{gol66}.

The discovery of planets in highly eccentric orbits, particularly in
multi-planet systems, presented a challenge for formation and
dynamical models. Significant progress has been made in the meantime
towards understanding these interesting systems (e.g.,
\citet{ras96,wei96,lin97}). For multi-planet systems, additional
constraints are imposed on eccentricity values due to the requirement
that the system remain dynamically stable. Dynamical interactions of
multi-planet systems in the context of eccentricity and formation
models have been investigated by numerous authors
\citep{for08,jur08,mal08,mat08,ray08,ray10,wan11,tim13}. From the
known multi-planet systems, there are four in particular which stand
out with respect to the diversity of eccentricities present within the
system. These are the HD~37605 \citep{coc04,wan12}, HD~74156
\citep{nae04,mes11}, HD~163607 \citep{gig12}, and HD~168443
\citep{mar99,mar01,pil11} systems. Each of these systems contain two
giant planets where the inner planet has an eccentricity larger than
0.5 whereas the orbit of the outer planet is much closer to
circular. It was shown by \citet{lau01} that dynamical interactions
between exoplanets can also result in subsequent radial velocity
variations that diverge from Keplerian orbital assumptions. It is
therefore interesting to consider the dynamical history and stability
of these systems as well as scenarios for their origins.

In this paper we present a detailed study of these four systems with
high eccentricity diversity. In Section 2 we summarize the orbital
configurations and system parameters for the systems discussed in this
study. In Section 3 we utilize the radial velocity solutions for these
systems to place quantitative limits on additional planets in each
system. Section 4 presents the dynamical analysis of each system
including the oscillations of the planet orbital eccentricities and
associated periastron precession. In Section 5 we discuss the effects
of mutual inclinations on the dynamical stabilities. Section 6
attempts to explore the origins of systems with such eccentricity
diversity based on our simulations. We discuss further implications of
these analyses in Section 7 and then provide concluding remarks in
Section 8.


\section{System Parameters}
\label{system}

\begin{figure*}
  \begin{center}
    \begin{tabular}{cc}
      \includegraphics[angle=270,width=7.5cm]{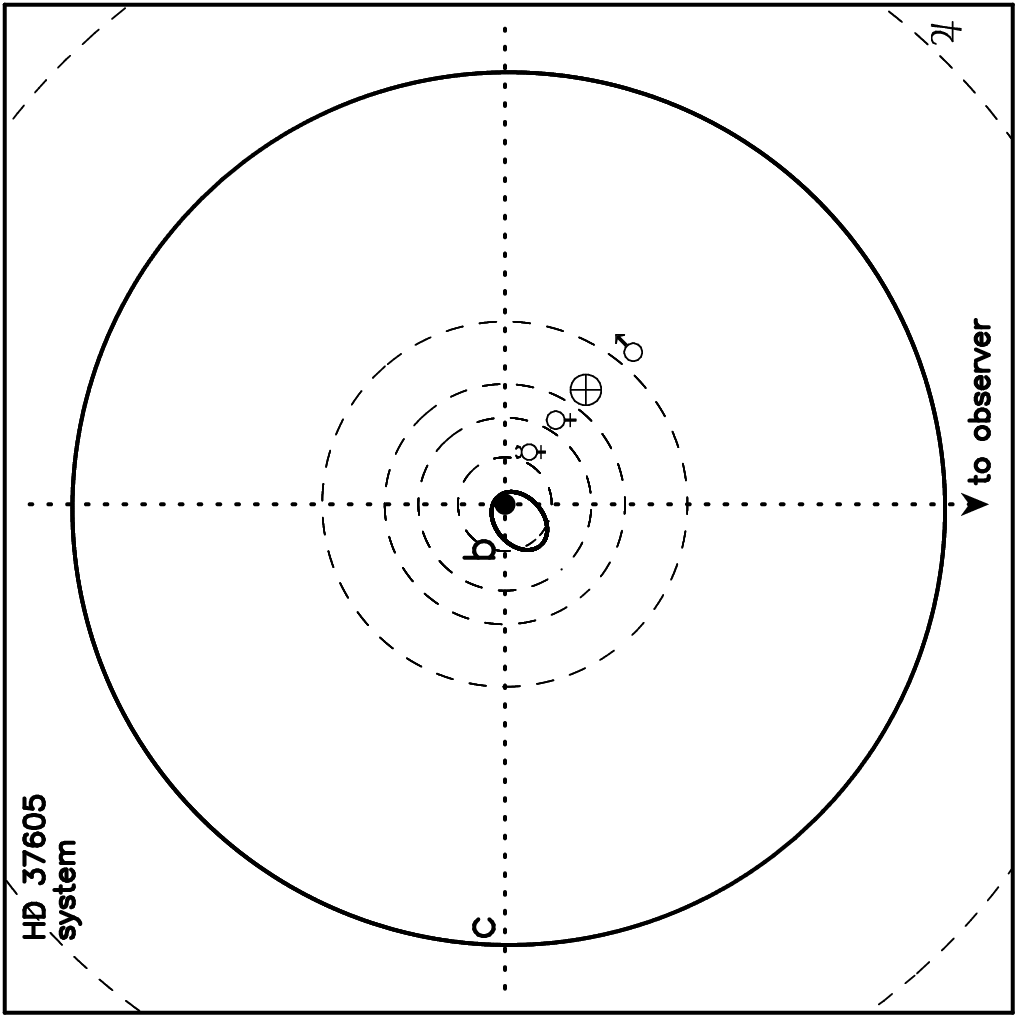} &
      \includegraphics[angle=270,width=7.5cm]{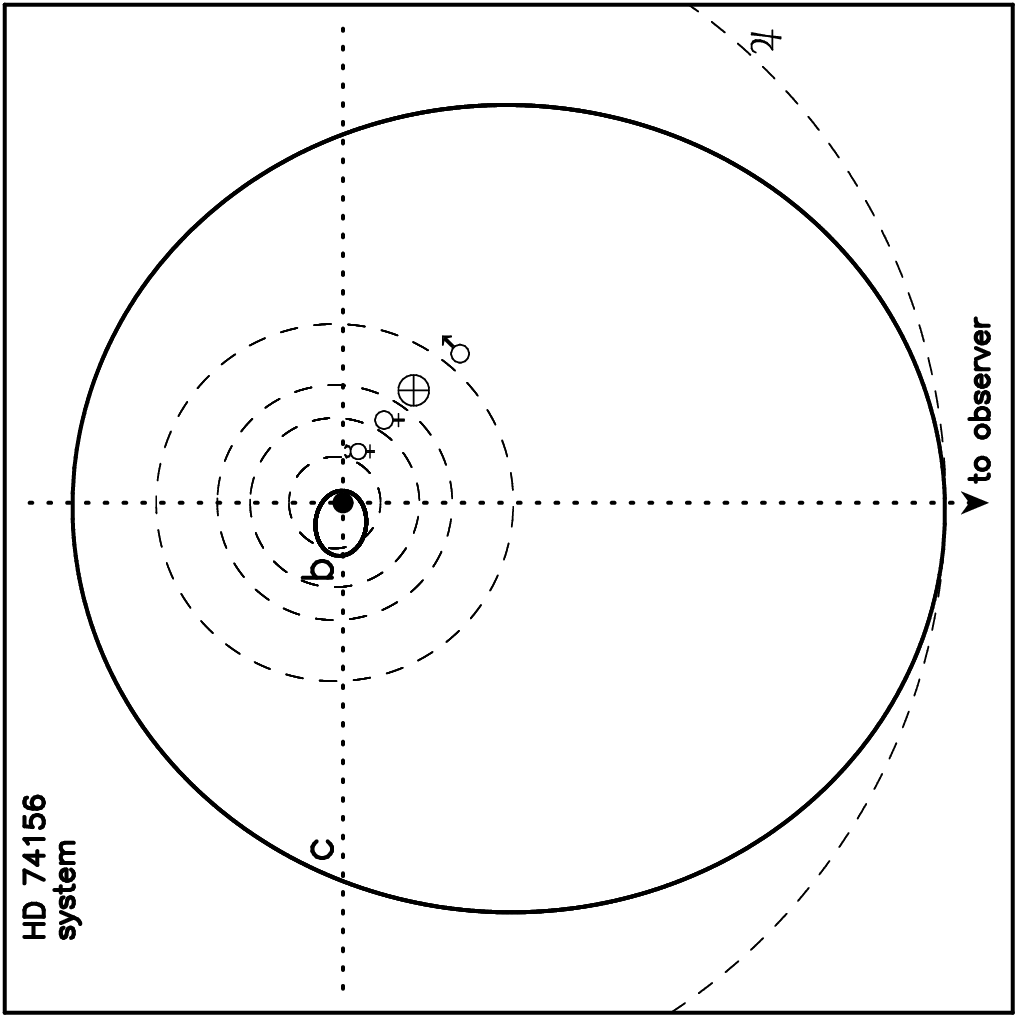} \\
      \includegraphics[angle=270,width=7.5cm]{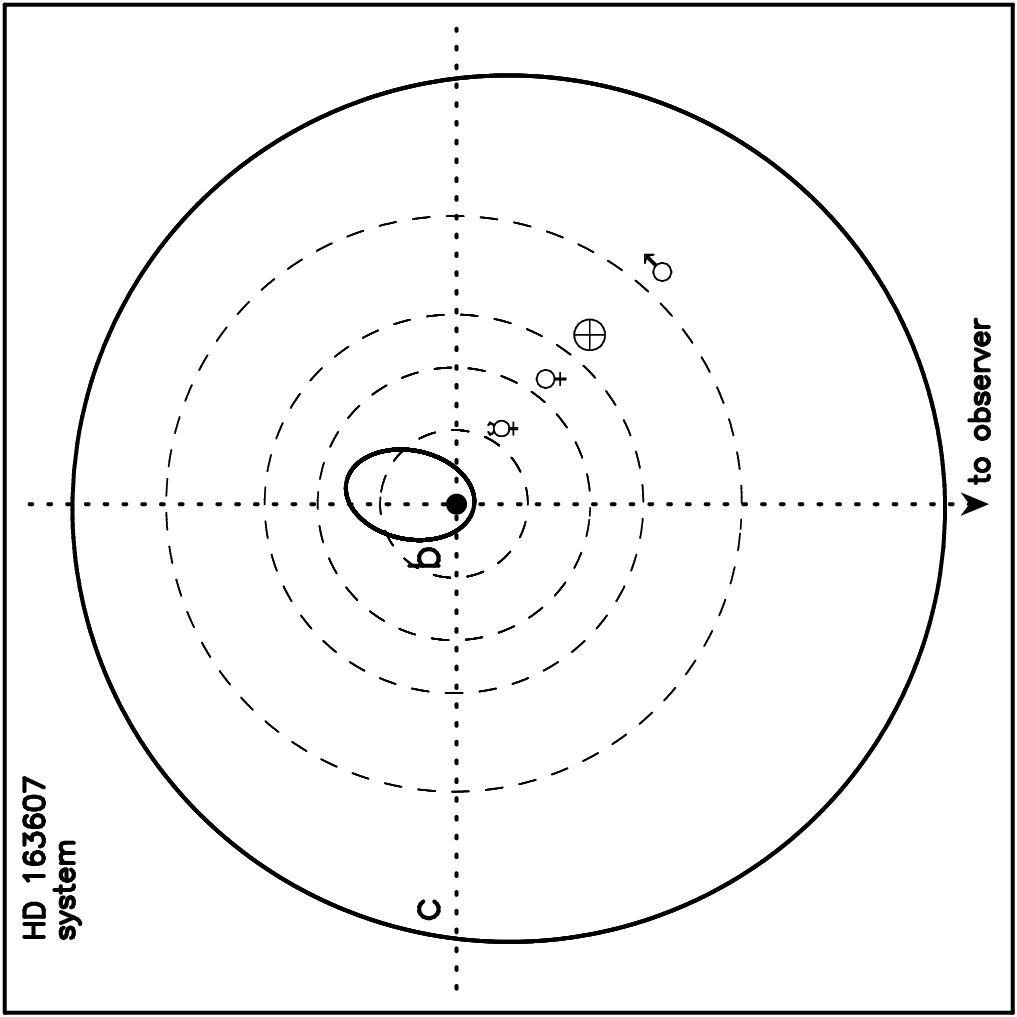} &
      \includegraphics[angle=270,width=7.5cm]{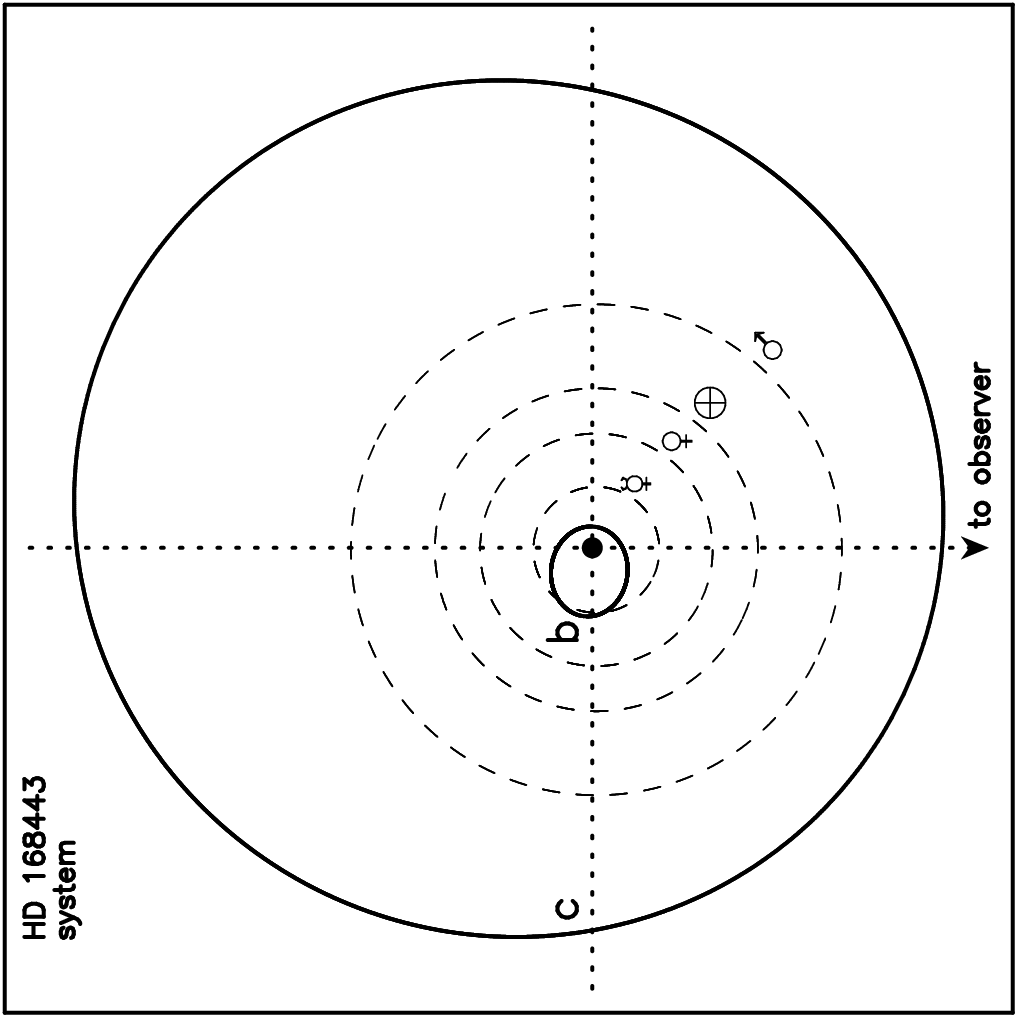}
    \end{tabular}
  \end{center}
  \caption{A top-down view of each system described in the paper:
    HD~37605 (top-left), HD~74156 (top-right), HD~163607
    (bottom-left), HD~168443 (bottom-right). The orbits of the system
    planets are shown as solid lines and the Solar System planets are
    included as dashed lines for comparison.}
  \label{orbits}
\end{figure*}

For exoplanetary systems in which there is significant orbital
eccentricity present, one must take care to prevent mis-interpretation
of the radial velocity data. There are several ambiguities which can
arise including confusing an eccentric orbit with 2:1 resonant systems
\citep{ang10}, 1:1 resonant co-orbital planets (``Trojan pairs'')
\citep{lau02,giu12}, and circular planets with long-period companions
\citep{rod09}. A concise summary of these confusion issues is
described by \citet{wit13} in which they test single planet eccentric
systems for multiplicity.

The analysis we present here focuses on four systems: HD~37605,
HD~74156, HD~163607, and HD~168443. A top-down view of each of these
systems is shown in Figure \ref{orbits} in which the orbits of both
system planets are shown (solid lines) along with orbits of the Solar
System planets for scale purposes (dashed lines). The criteria for
selecting these systems was twofold. Firstly, the system was required
to be a multi-planet system in which the inner planet has an
eccentricity greater than 0.5. Secondly, the system must have
sufficient data, including at least one complete orbital phase of the
outer planet, to minimize the potential confusion issues mentioned
above.

\begin{deluxetable*}{lcccc}
  \tablecolumns{5}
  \tablewidth{0pc}
  \tablecaption{\label{paramtab} System Parameters}
  \tablehead{
    \colhead{Parameter} &
    \multicolumn{2}{c}{HD~37605$^{(1)}$} &
    \multicolumn{2}{c}{HD~74156$^{(2)}$} \\
    \colhead{} &
    \colhead{b} &
    \colhead{c} &
    \colhead{b} &
    \colhead{c}
  }
  \startdata
$M_\star$ ($M_\odot$)
  & \multicolumn{2}{c}{$1.000 \pm 0.050$}
  & \multicolumn{2}{c}{$1.24$} \\
$P$ (days)
  & $55.01307 \pm 0.00064$ & $2720 \pm 57$
  & $51.638 \pm 0.004$     & $2520 \pm 15$ \\
$T_p\,^{(5)}$
  & $13378.241 \pm 0.020$  & $14838 \pm 581$
  & $10793.3 \pm 0.2$      & $8416 \pm 33$ \\
$e$
  & $0.6767 \pm 0.0019$    & $0.013 \pm 0.015$
  & $0.63 \pm 0.01$        & $0.38 \pm 0.02$ \\
$\omega$ (deg)
  & $220.86 \pm 0.28$      & $221 \pm 78$
  & $174 \pm 2$            & $268 \pm 4$ \\
$K$ (m\,s$^{-1}$)
  & $202.99 \pm 0.72$      & $48.90 \pm 0.86$
  & $108 \pm 4$            & $115 \pm 3$ \\
$M_p \sin i$ ($M_J$)
  & $2.802 \pm 0.011$      & $3.366 \pm 0.072$
  & $1.78 \pm 0.04$        & $8.2 \pm 0.2$ \\
$a$ (AU)
  & $0.2831 \pm 0.0016$    & $3.814 \pm 0.058$
  & $0.29169 \pm 0.00001$  & $3.90 \pm 0.02$ \\
\hline
\hline
\rule{0pt}{3ex}
Parameter & \multicolumn{2}{c}{HD~163607$^{(3)}$}
& \multicolumn{2}{c}{HD~168443$^{(4)}$} \\
 & b & c & b & c \\
\hline
\rule{0pt}{3ex}
$M_\star$ ($M_\odot$)
  & \multicolumn{2}{c}{$1.09 \pm 0.02$}
  & \multicolumn{2}{c}{$0.995 \pm 0.019$} \\
$P$ (days)
  & $75.29 \pm 0.02$       & $1314 \pm 8$
  & $58.11247 \pm 0.0003$  & $1749.83 \pm 0.57$ \\
$T_p\,^{(5)}$
  & $14185.00 \pm 0.24$    & $15085 \pm 880$
  & $15626.199 \pm 0.024$  & $15521.3 \pm 2.2$ \\
$e$
  & $0.73 \pm 0.02$        & $0.12 \pm 0.06$
  & $0.52883 \pm 0.00103$  & $0.2113 \pm 0.00171$ \\
$\omega$ (deg)
  & $78.7 \pm 2.0$         & $265 \pm 93$
  & $172.923 \pm 0.139$    & $64.87 \pm 0.5113$ \\
$K$ (m\,s$^{-1}$)
  & $51.1 \pm 1.4$         & $40.4 \pm 1.3$
  & $475.133 \pm 0.9102$   & $297.70 \pm 0.618$ \\
$M_p \sin i$ ($M_J$)
  & $0.77 \pm 0.04$        & $2.29 \pm 0.16$
  & $7.659 \pm 0.0975$     & $17.193 \pm 0.21$ \\
$a$ (AU)
  & $0.36 \pm 0.01$        & $2.42 \pm 0.01$
  & $0.2931 \pm 0.00181$   & $ 2.8373 \pm 0.018$
  \enddata
  \tablenotetext{(1)}{\citet{wan12}}
  \tablenotetext{(2)}{\citet{mes11}}
  \tablenotetext{(3)}{\citet{gig12}}
  \tablenotetext{(4)}{\citet{pil11}}
  \tablenotetext{(5)}{JD -- 2,440,000}
\end{deluxetable*}

HD~37605b was the first planet discovered using the Hobby-Eberly
Telescope \citep{coc04}. Further studies by \citet{wan12} refined the
orbit of the inner planet as well as detecting an additional planet in
a long-period orbit. The HD~74156 system was discovered by
\citet{nae04} using ELODIE data. The orbits of the both planets were
further refined by \citep{mes11} using Keck/HIRES data. The HD~163607
system was another case in which both planets were announced
simultaneously, this time by \citet{gig12} with Keck/HIRES
data. HD~168443b was discovered by \citet{mar99}, who also detected
evidence of a trend due to an additional planet. This trend was later
confirmed to be a second planet by \citet{mar01}. The orbits of both
planets were later refined by \citet{pil11}.

\begin{figure}
  \includegraphics[angle=270,width=8.2cm]{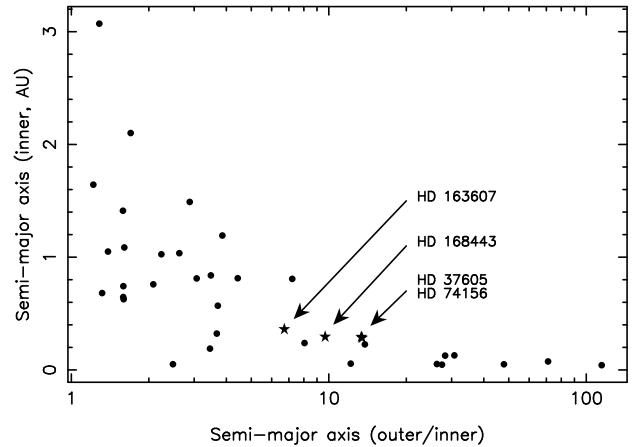}
  \caption{This plot includes all 2-planet systems detected using the
    radial velocity technique. The semi-major axis of the inner planet
    is plotted as a function of the ratio of the outer to inner planet
    semi-major axes. The locations of the four systems studied in this
    paper are shown as stars.}
  \label{comp}
\end{figure}

We summarize the Keplerian orbital solutions for each of these systems
in Table \ref{paramtab}. These are the parameters used throughout the
remainder of the paper to perform analyses of the orbits. One striking
feature seen in Figure \ref{orbits} and Table \ref{paramtab} is that
these particular kinds of systems are all approximately the same size
scale. This is not a selection effect nor due to observational bias
since the detection of the outer planet resulted from continued
monitoring. The semi-major axes of the planets range between
0.28--0.36~AU and 2.4--3.9~AU for the inner and outer planets
respectively. To quantify the similarity of the system size scales, we
compared these systems with others by extracting orbital data for all
2-planet systems detected using the radial velocity method from the
Exoplanet Data Explorer\footnote{\tt http://exoplanets.org/}
\citep{wri11}. The data are current as of 27th September 2013. Figure
\ref{comp} plots the semi-major axes of the inner planet as a function
of the ratio of outer to inner planet semi-major axes. For the four
systems considered here, these ratios are 13.47, 13.37, 6.72, 9.68 for
HD~37605, HD~74156, HD~163607, and HD~168443 respectively (depicted as
stars in Figure \ref{comp}). Note that HD~37605 and HD~74156 are
almost identical in this respect and so are almost indistinguishable
in Figure \ref{comp}. All four systems occupy a relatively small
fraction of the known distribution shown in this figure, possibly due
to the dynamical constraints imposed by the eccentricity of the inner
planet.

\section{Limits on Additional Planets}
\label{limits}

Before we investigated the dynamical interactions of the 2-planet
systems, we first tested both the Keplerian orbital solutions
previously found (see Table \ref{paramtab}) and determined if the
residuals may disguise the presence of further planetary
companions. The radial velocity data were obtained from the published
literature and from the NASA Exoplanet Archive\footnote{\tt
  http://exoplanetarchive.ipac.caltech.edu/} \citep{ake13}. We fit the
radial velocity data using the partially linearized, least-squares
fitting procedure described in \citet{wri09} and estimated parameter
uncertainties using the BOOTTRAN bootstrapping routines described in
\citet{wan12}. We obtained Keplerian orbital solutions consistent with
those described in Table \ref{paramtab} and thus adopt the rms
residuals of those analyses.

\begin{figure*}
  \begin{center}
    \begin{tabular}{cc}
      \includegraphics[angle=270,width=8.2cm]{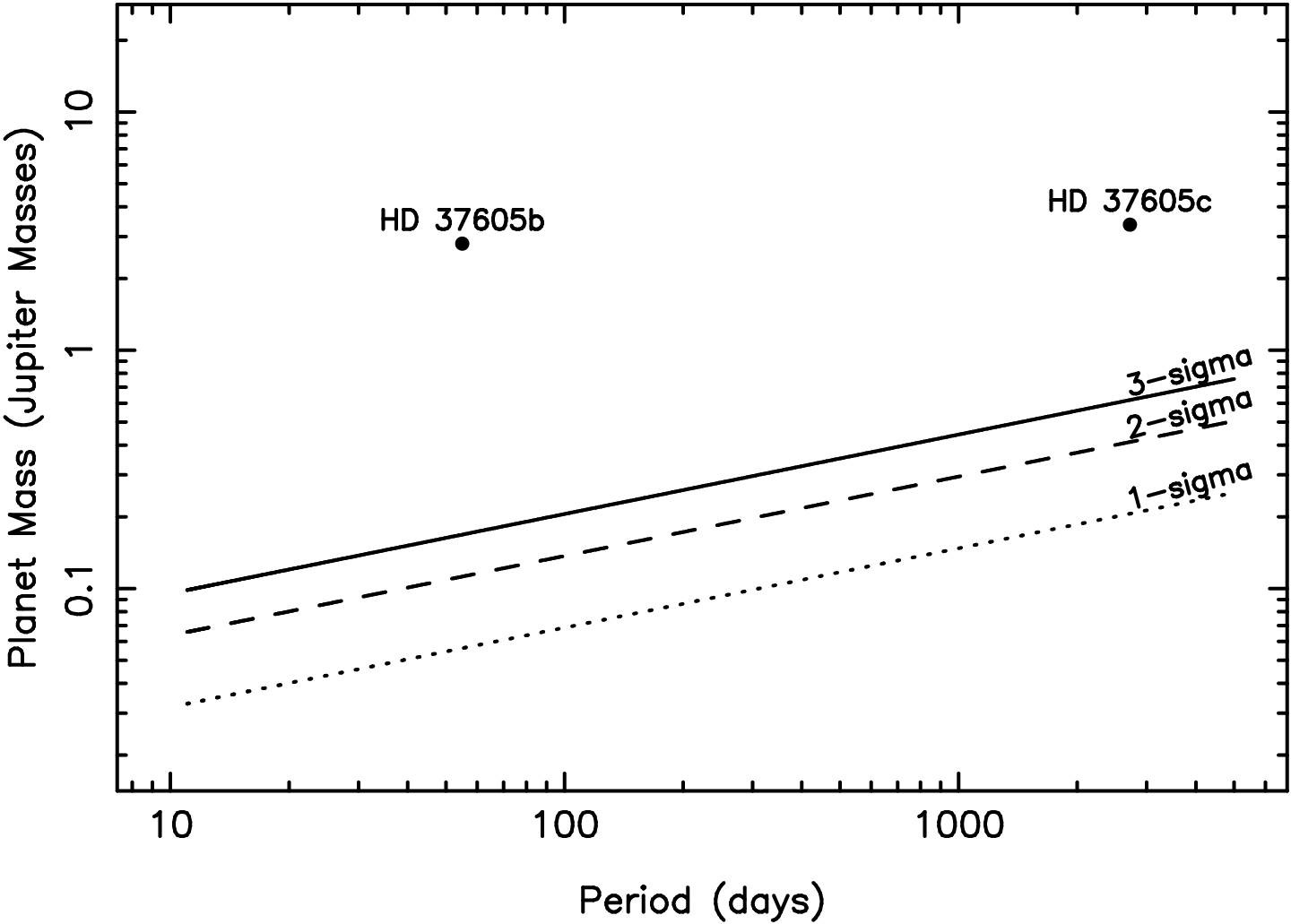} &
      \includegraphics[angle=270,width=8.2cm]{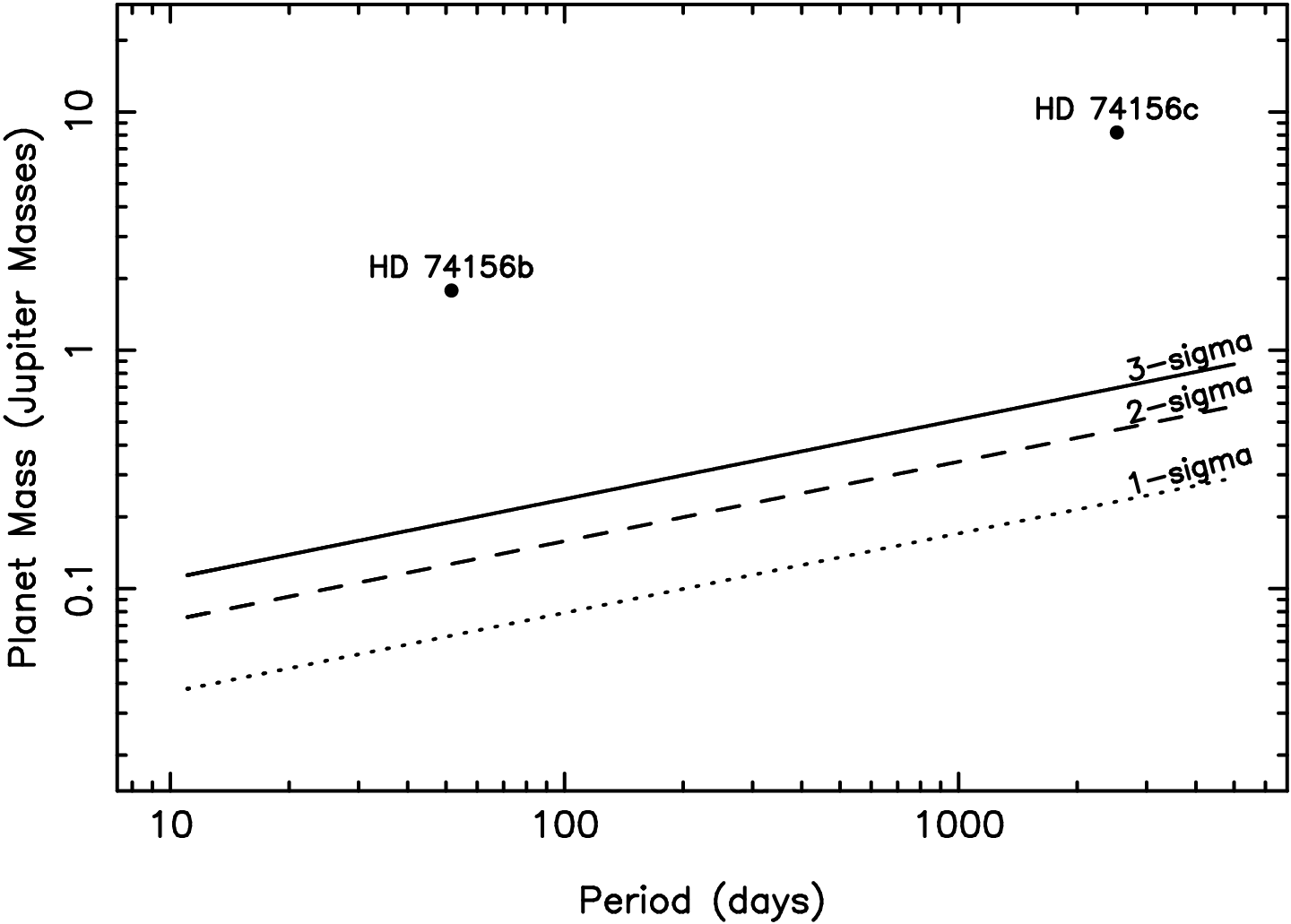} \\
      \includegraphics[angle=270,width=8.2cm]{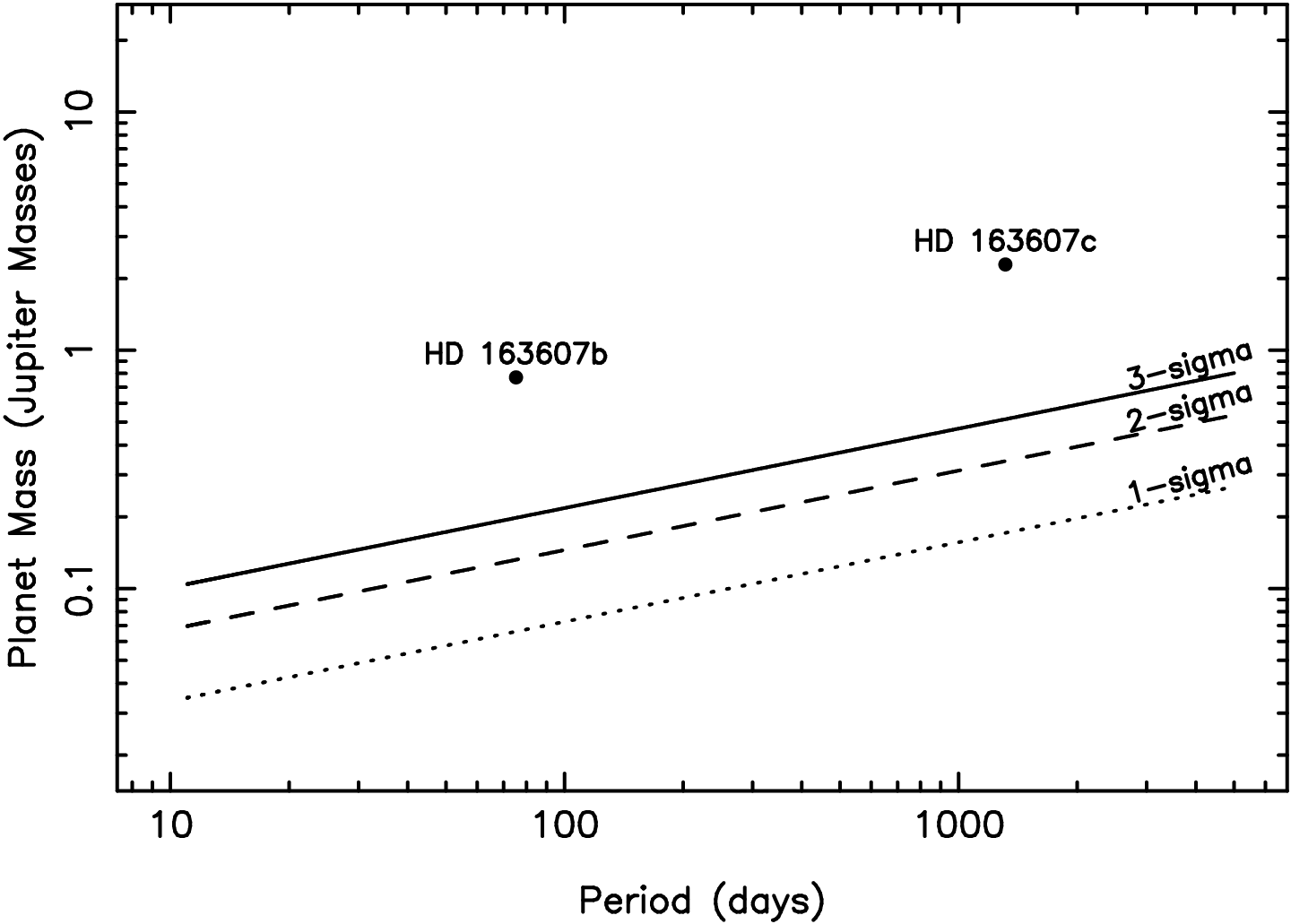} &
      \includegraphics[angle=270,width=8.2cm]{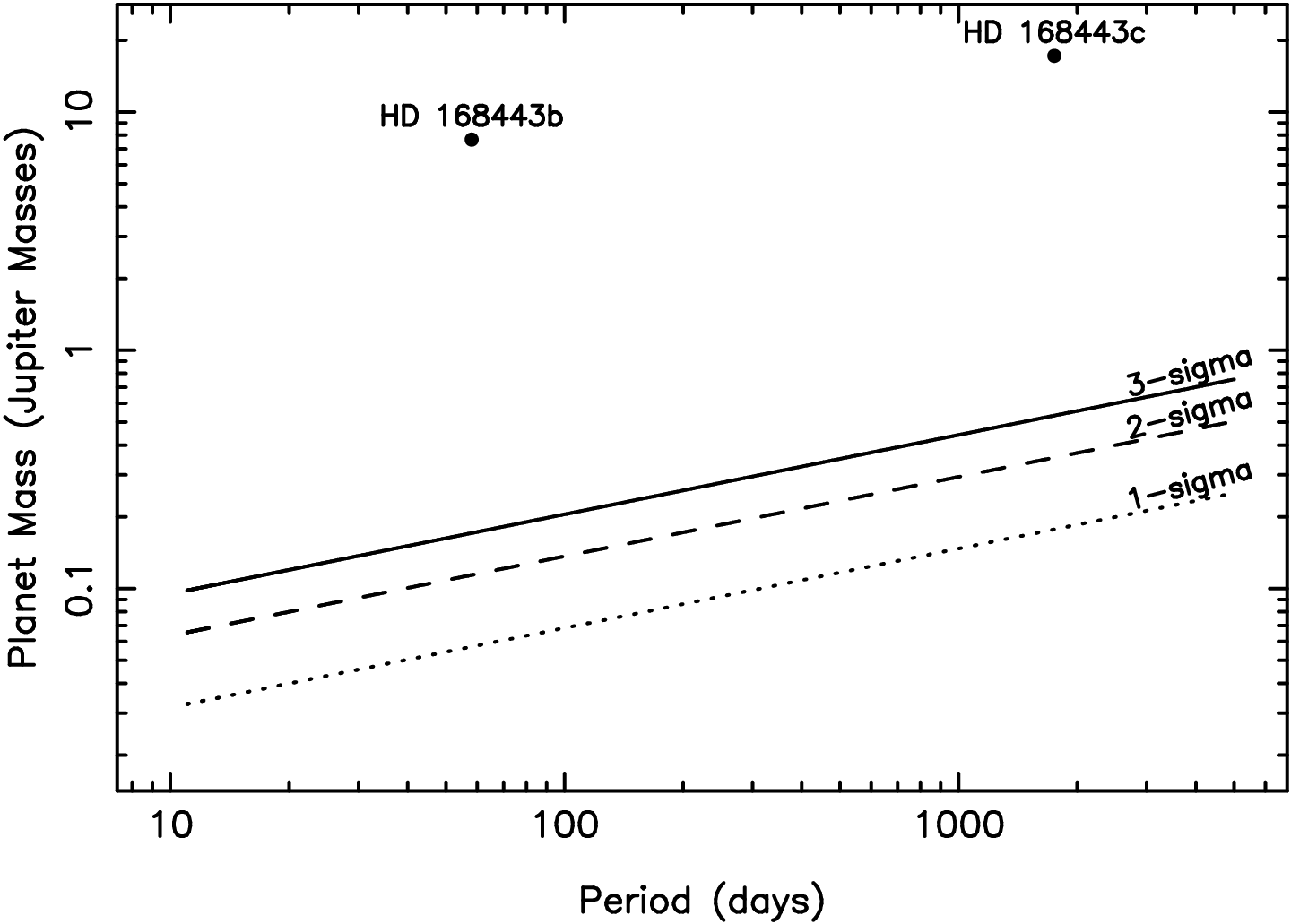} \\
    \end{tabular}
  \end{center}
  \caption{Exclusion regions (above the lines) for additional planets
    with the HD~37605 (top-left), HD~74156 (top-right), HD~163607
    (bottom-left), and HD~168443 (bottom-right) systems. The
    1--3~$\sigma$ exclusion boundaries are based upon the rms scatter
    of the residuals to radial velocity data after the best-fit
    Keplerian model has been removed. The detected planets in each
    system are shown for reference.}
  \label{limitsfig}
\end{figure*}

The rms residuals are used to place 1--3~$\sigma$ thresholds which
limits the presence of additional planetary companions in each system
as a function of both planetary mass and orbital period. This is
achieved using the maximum semi-amplitude of the radial velocity
allowed by the residuals. This is expressed as
\begin{equation}
  K = \left( \frac{2 \pi G}{P} \right)^{1/3} \frac{M_p \sin
    i}{(M_\star + M_p)^{2/3}} \frac{1}{\sqrt{1-e^2}}
\end{equation}
where $P$ is the period, $i$ is the inclination of the planetary orbit,
$e$ is the eccentricity, and $M_p$ and $M_\star$ are the masses of the
planet and parent star respectively. We assume a circular orbits which
is a reasonable assumption for all but high eccentricities which would
likely render the system unstable. The resulting mass limit thresholds
are shown for each system in Figure \ref{limitsfig}. In each case the
3$\sigma$ threshold lies substantially below the mass range of the two
known planets. Due to the similarity in both the amplitude of the
residuals and the mass of the system host stars, the constraints on
possible undetected planetary masses as a function of orbital period
is almost identical for all four systems. The similarity of the
systems is further emphasized by the use of identical axis scales on
each of the plots.

For each system, there is usually a combination of radial velocity
data sources which often result in the rms scatter of the fit
residuals being dominated by one or more of those sources. In these
situations we adopt the rms scatter for the highest quality data to
establish the limits of additional planets described here. For
HD~37605, the rms scatter of the residuals is 7.61~m\,s$^{-1}$ when
HET data are included but the Keck/HIRES data alone produce residuals
of 2.08~m\,s$^{-1}$. For HD~74156, the rms scatter of 12.8~m\,s$^{-1}$
is dominated by CORALIE and ELODIE data. The rms scatter of the Keck
data residuals alone is 3.5~m\,s$^{-1}$. The radial velocity data for
both HD~163607 and HD~168443 were obtained exclusively from Keck/HIRES
and the resulting rms scatter of the residuals is 2.9~m\,s$^{-1}$ and
3.9~m\,s$^{-1}$ respectively.

\begin{deluxetable}{lcc}
  \tablecolumns{5}
  \tablewidth{0pc}
  \tablecaption{\label{limitstab} Planetary Upper Mass Limits}
  \tablehead{
    \colhead{System} &
    \multicolumn{2}{c}{3$\sigma$ mass limit ($M_J$)} \\
    \colhead{} &
    \colhead{$P = 100$~days} &
    \colhead{$P = 1000$~days}
  }
  \startdata
  HD~37605  & 0.20 & 0.44 \\
  HD~74156  & 0.24 & 0.51 \\
  HD~163607 & 0.22 & 0.47 \\
  HD~168443 & 0.20 & 0.44
  \enddata
\end{deluxetable}

\begin{figure*}
  \begin{center}
    \includegraphics[angle=270,width=15.0cm]{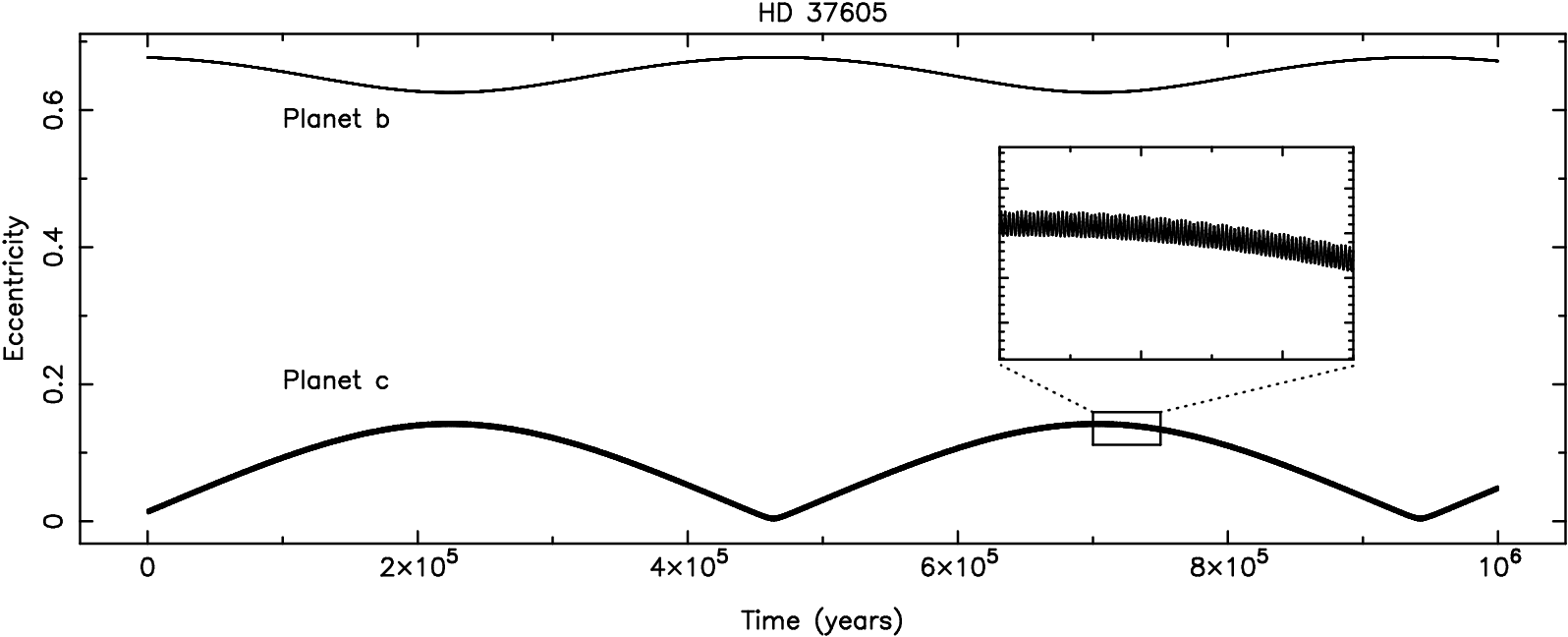} \\
    \includegraphics[angle=270,width=15.0cm]{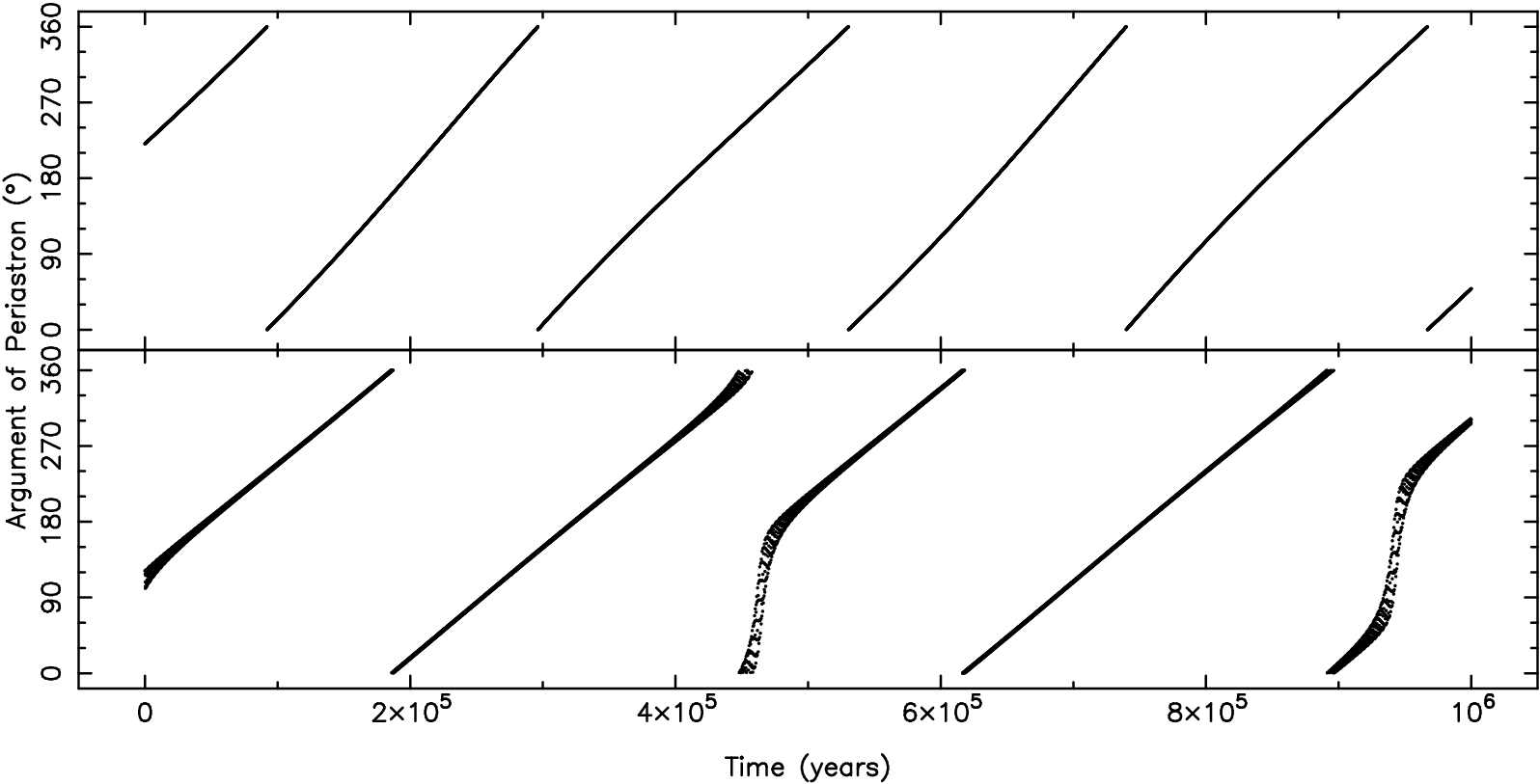}
  \end{center}
  \caption{Dynamical simulations of the HD~37605 system, showing the
    eccentricity oscillations of both planets (top panel) and the
    periastron precession of the b planet (middle panel) and c planet
    (bottom panel). The zoom window in the top panel shows a
    simulation period of 50,000 years.}
  \label{hd37605fig}
\end{figure*}

The planetary upper limits imposed by this analysis for each system
are shown in Table \ref{limitstab} at orbital periods of 100 and 1000
days. This shows similar results of planets more massive than
0.2--0.5~$M_J$ being excluded from lying between the known planets at
the 3$\sigma$ level. This is likely due to instability regions imposed
by the oscillating eccentricities of the much more massive inner and
outer planet. We explore these oscillations in the following section.


\section{Dynamical Analysis}
\label{dynamics}

The stability of exoplanetary systems has been explored in
considerable depth by such authors as \citet{cha96,bar06b,bar07}. In
this section we examine the dynamical interactions of the four
particular two-planet systems which are the subject of this paper. The
N-body integrations required for the dynamical simulations were
performed by utilizing the Mercury Integrator Package, described in
more detail by \citet{cha99}. For these simulations, we adopted the
hybrid symplectic/Bulirsch-Stoer integrator. We also used a Jacobi
coordinate system which generally provides more accurate results for
multi-planet systems \citep{wis91,wis06} except in cases of close
encounters \citep{cha99}. The integrations were performed for a
simulation of $10^6$ years, in steps of 100 years, starting at the
present epoch.


\subsection{The HD~37605 System}
\label{hd37605}

The initial eccentricity and argument of periastron for the HD~37605 b
and c planets are 0.6767, 220\degr and 0.013, 221\degr respectively
(see Table \ref{paramtab}). The results of the N-body integrations are
plotted in Figure \ref{hd37605fig}. The eccentricity oscillations
shown in the top panel complete approximately two cycles during the
$10^6$~year simulation. The range of eccentricity for the b and c
planets are 0.626--0.677 and 0.001--0.145 respectively. Thus, in this
case, the amplitude of the eccentricity variations for the outer
planet exceeds that of the inner planet.

\begin{figure*}
  \begin{center}
    \includegraphics[angle=270,width=15.0cm]{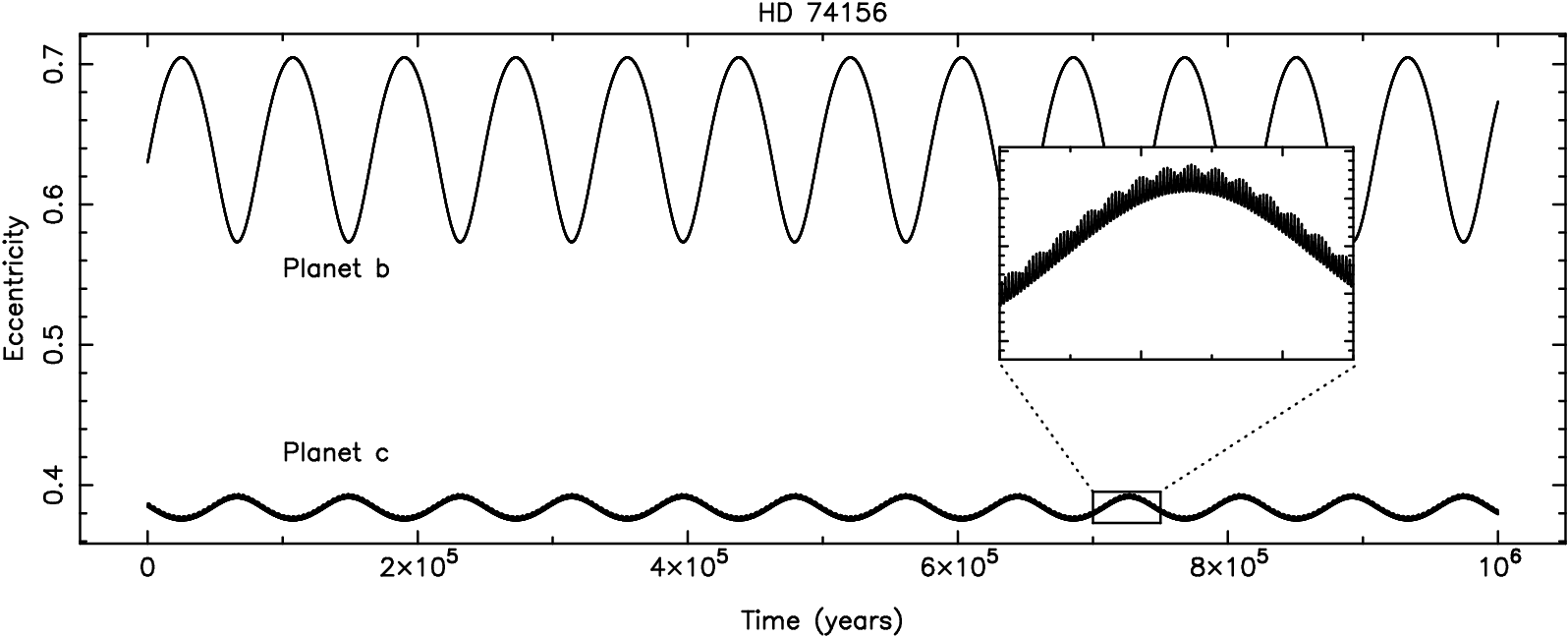} \\
    \includegraphics[angle=270,width=15.0cm]{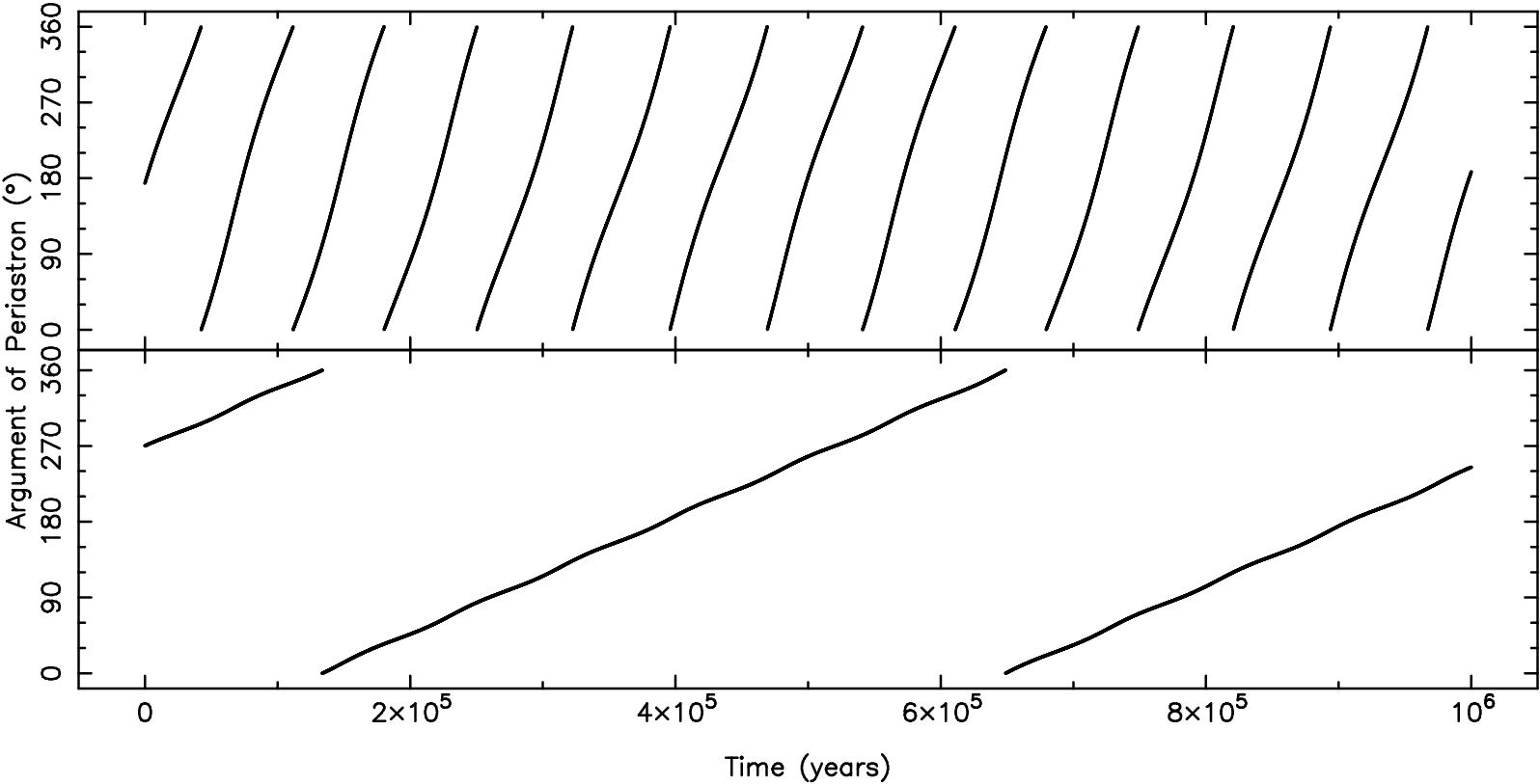}
  \end{center}
  \caption{Dynamical simulations of the HD~74156 system, showing the
    eccentricity oscillations of both planets (top panel) and the
    periastron precession of the b planet (middle panel) and c planet
    (bottom panel). The zoom window in the top panel shows a
    simulation period of 50,000 years.}
  \label{hd74156fig}
\end{figure*}

These results have several details of note. Firstly, the
high-amplitude eccentricity oscillations of the outer planet are
comprised of smaller-amplitude higher-frequency oscillations. This is
shown by the zoomed-in region in the top panel of Figure
\ref{hd37605fig} which has a time-span of 50,000 years. The amplitude
of these higher-frequency oscillations is $\sim 0.05$ with a period of
$\sim 550$ years. Secondly, the periastron arguments of the planets
begin closely aligned (see Table \ref{paramtab}) but differing
precession rates result in them being slightly out of sync with each
other. The precession of the inner planet (0.164\degr/century) is
slightly higher than that for the outer planet
(0.14\degr/century). Note also that during those moments when the
eccentricity of the outer planet is $\sim 0.0$ then the periastron
argument becomes highly uncertain (since $\omega$ is undefined when $e
= 0.0$) resulting in a dispersion of the $\omega$ values near those
times. For this reason the evolution of the periastron argument for
the outer planet is difficult to determine and may possibly match that
of the inner planet.


\subsection{The HD~74156 System}
\label{hd74156}

The HD~74156 system differs from the other three in this study in that
both the mass ratio of the two planets and the eccentricity of the
outer planet are the highest of all four systems. As shown in Figure
\ref{orbits} and Table \ref{paramtab}, the orientation of the orbits
are not closely aligned as was the case for HD~37605.  The N-body
integration results for HD~74156 are plotted in Figure
\ref{hd74156fig}. The eccentricity oscillations for both planets shown
in the top panel complete approximately 12 cycles during the
$10^6$~year simulation. The range of eccentricity for the b and c
planets are 0.573--0.705 and 0.375--0.394 respectively. The major
eccentricity variation in this case occurs for the inner planet
whereas the variations for the outer planet are far more subtle. The
high-frequency oscillations shown in the zoom-in window of the top
panel of Figure \ref{hd74156fig} show evidence of vibrational beating
with an amplitude that is negligible compared with the low-frequency
oscillations. Although the first-order eccentricity oscillations of
the planets share the same frequency, the periastron precessions
(shown in the bottom two panels) do not. The precession rates of the
inner and outer planet are 0.05\degr/century and 0.007\degr/century
respectively.


\subsection{The HD~163607 System}
\label{hd163607}

\begin{figure*}
  \begin{center}
    \includegraphics[angle=270,width=15.0cm]{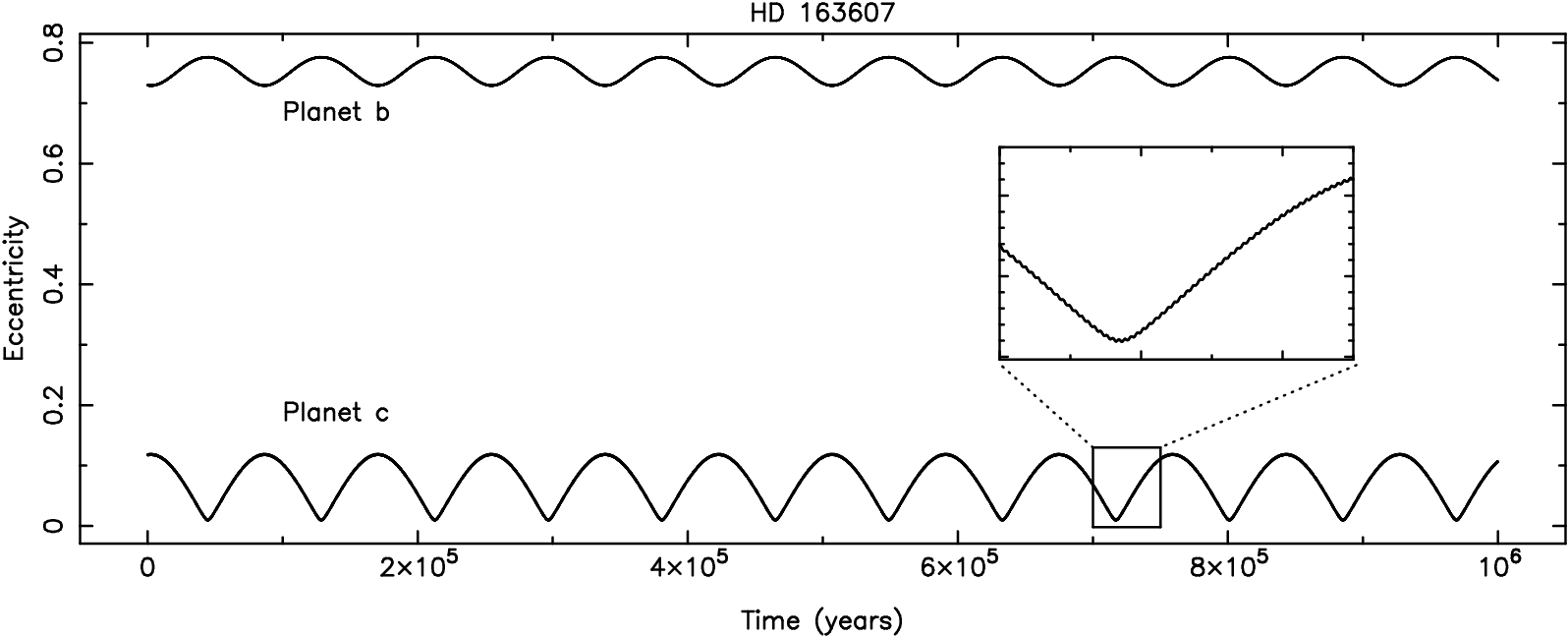} \\
    \includegraphics[angle=270,width=15.0cm]{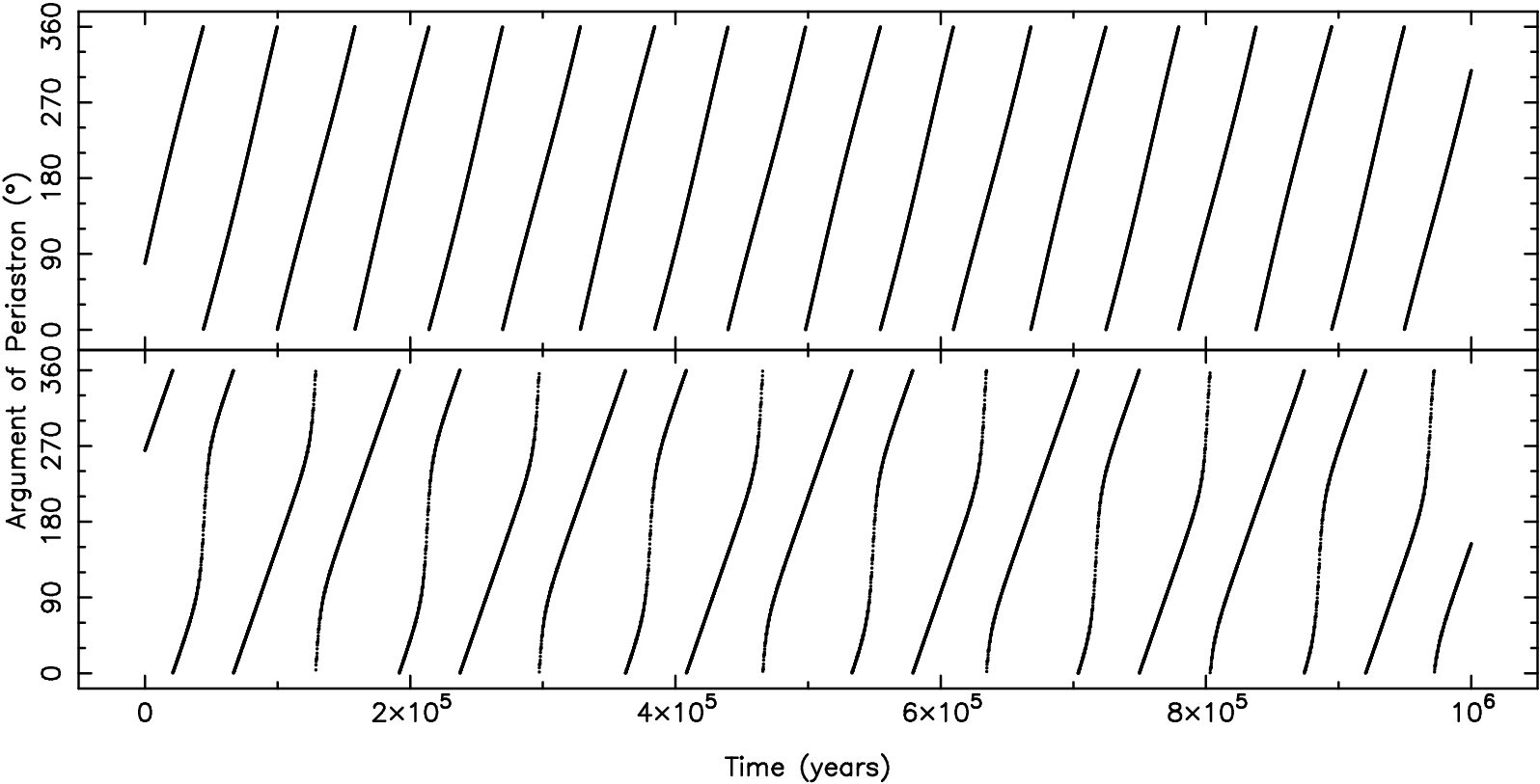}
  \end{center}
  \caption{Dynamical simulations of the HD~163607 system, showing the
    eccentricity oscillations of both planets (top panel) and the
    periastron precession of the b planet (middle panel) and c planet
    (bottom panel). The zoom window in the top panel shows a
    simulation period of 50,000 years.}
  \label{hd163607fig}
\end{figure*}

Of the four systems, the HD~163607 system contains both the largest
orbital period and the highest eccentricity for the inner
planet. Additionally, the periastron arguments for the inner and outer
planets are almost perfectly anti-aligned ($\pi$ out of phase) with
one another. The N-body integration results for HD~163607 are plotted
in Figure \ref{hd163607fig}. This figure shows that both the
oscillation of the eccentricities and periastron precession of the
orbits (0.06\degr/century) remain approximately in sync for the two
planets during the 100,000 year simulation. Note also that the outer
planet also exhibits higher-frequency eccentricity oscillations
although at a much smaller amplitude than for the outer planet in the
other three systems (see the zoom-in window in the top panel of Figure
\ref{hd163607fig}. The range of eccentricity for the b and c planets
are 0.729--0.776 and 0.009--0.119 respectively. Thus, the major
oscillations occur for the outer planet. The eccentricity of the outer
planet is periodically forced to zero which produces a similar
ambiguity in the periastron argument as was seen in the case of
HD~37605 (see Section \ref{hd37605}).


\subsection{The HD~168443 System}
\label{hd168443}

\begin{figure*}
  \begin{center}
    \includegraphics[angle=270,width=15.0cm]{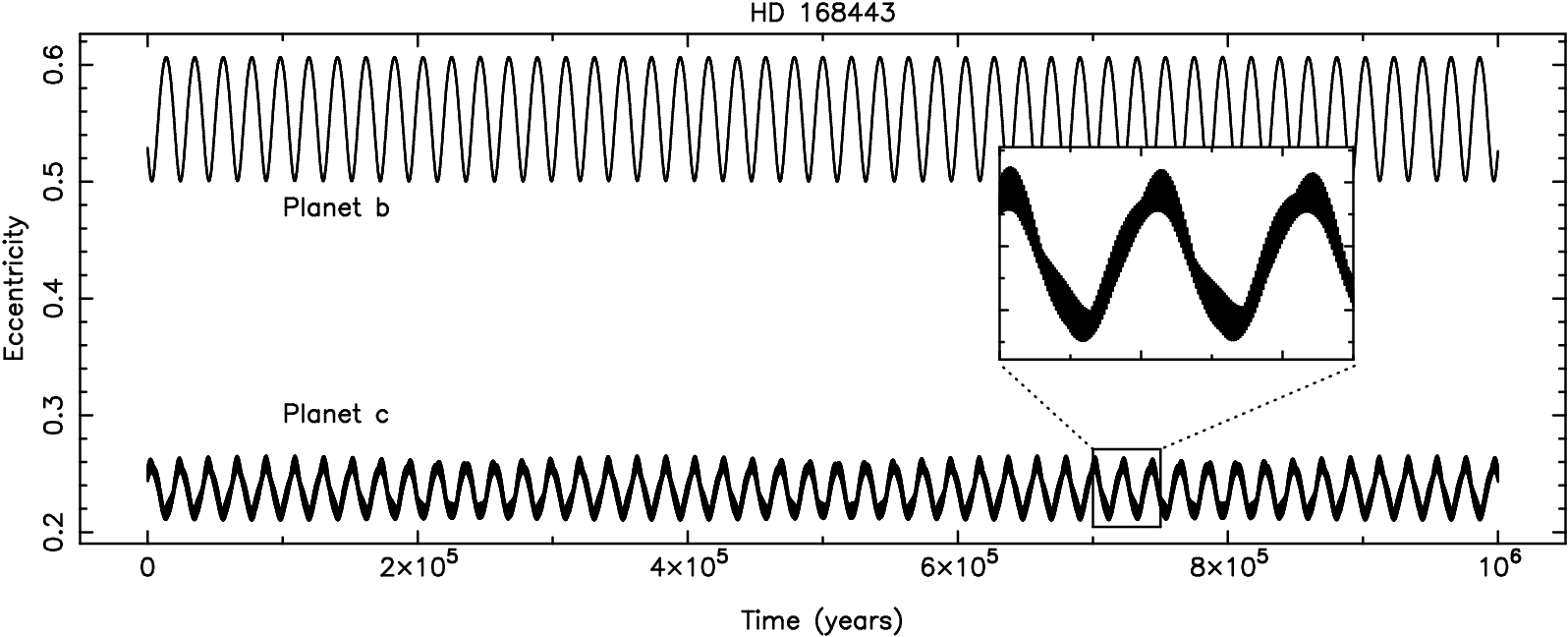} \\
    \includegraphics[angle=270,width=15.0cm]{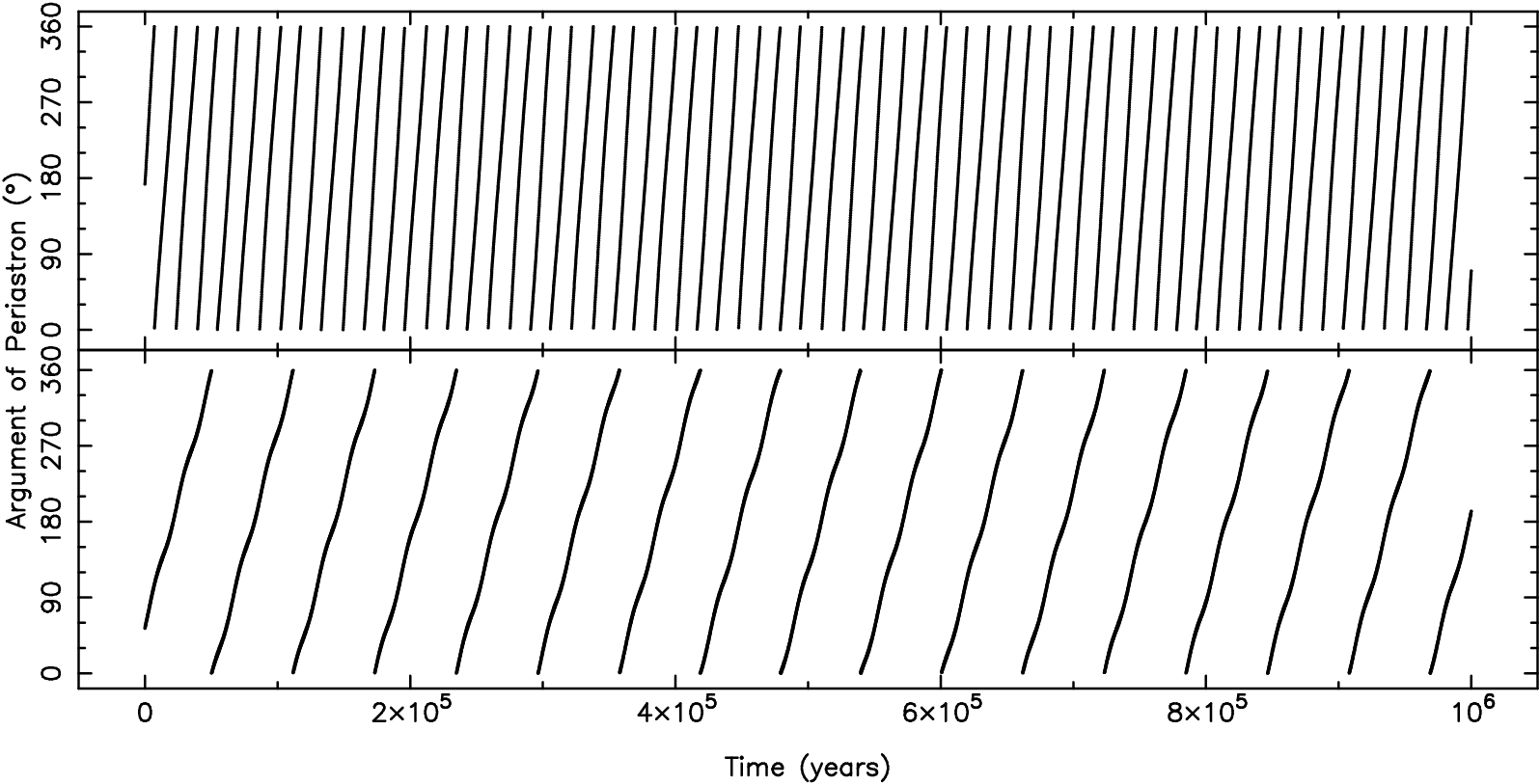}
  \end{center}
  \caption{Dynamical simulations of the HD~168443 system, showing the
    eccentricity oscillations of both planets (top panel) and the
    periastron precession of the b planet (middle panel) and c planet
    (bottom panel). The zoom window in the top panel shows a
    simulation period of 50,000 years.}
  \label{hd168443fig}
\end{figure*}

The stability of the HD~168443 system has been previously studied by
\citet{bar04} in which they found that the system is stable although
weak interactions occur between the two planets. The revised orbital
parameters provided by \citet{pil11} do not result in a substantial
change to the conclusion of system stability. The planets of the
HD~168443 system have the largest minimum masses of all four
systems. This results in substantially increased interactions between
the two planets and a relatively high-frequency of both the
eccentricity oscillations and periastron precession of the orbits. The
results of the N-body integration for HD~168443 are shown in Figure
\ref{hd168443fig}. The range of eccentricity for the b and c planets
are 0.500--0.607 and 0.210--0.265 respectively. Although the
eccentricity oscillations have almost the same frequency, the
amplitude of the outer planet oscillations are almost half that of the
inner planet and have a high-frequency component as seen in the other
three systems. The periastron precession rate of the inner planet is
substantial: 0.23\degr/century compared with 0.06\degr/century for the
outer planet.

\section{Mutual Inclinations}
\label{mutualinc}

\begin{figure*}
  \begin{center}
    \includegraphics[angle=270,width=15.0cm]{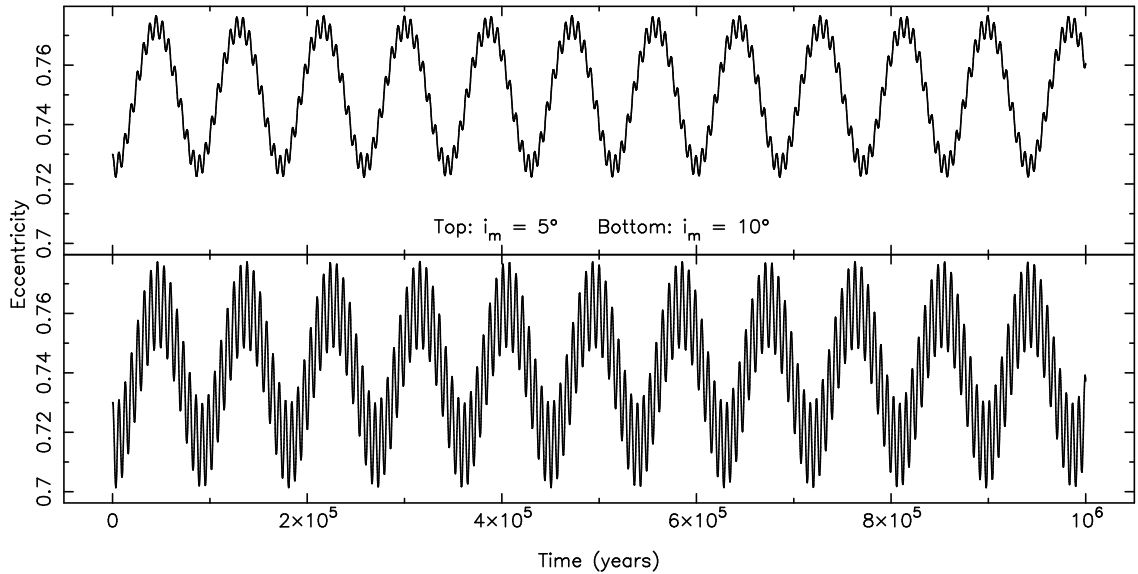}
  \end{center}
  \caption{The effect of introducing mutual inclination ($i_m$) to the
    HD~163607 system on the eccentricity evolution of planet b. Top
    panel: $i_m = 5$\degr. Bottom panel: $i_m = 10$\degr.}
  \label{mutualincfig}
\end{figure*}

Here we briefly investigate the effect of introducing a mutual
inclination into the results of our simulations and the predicted
angular momentum exchange. We performed this investigation for the
HD~163607 system since the results for this system shown in Section
\ref{dynamics} yield the most well-defined eccentricity oscillations
of the four systems considered. To do this, we repeated the N-body
integrations with mutual orbital inclinations from 0\degr to 90\degr
in increments of 5\degr. Shown in Figure \ref{mutualincfig} are the
results for 5\degr and 10\degr mutual inclinations where the
eccentricity evolution of the b planet is shown for the full 100,000
year simulation.

The effect of introducing a mutual inclination is to increase the
amplitude of both the primary (low-frequency) and secondary
(high-frequency) oscillations. This increase in oscillation amplitude
increases with increasing mutual inclination until the inner planet is
ejected from the system, the timescale of which depends on the amount
of mutual inclination present. For example, the HD~163607 system does
not remain stable for the full 100,000 year simulation for mutual
inclinations larger than 60\degr. Although the amplitude of the
secondary oscillations increases with increasing mutual inclination
(see Figure \ref{mutualincfig}), the frequency of the oscillations
remains the same. Thus it would require several hundred years of
observations (in the case of HD~163607) in order to detect the extent
of the mutual inclination from such eccentricity variations.

\section{Origin of Systems with Unequal Eccentricities}
\label{origin}

We have attempted to understand the origin of these systems with large
differences in eccentricity. The large eccentricities of exoplanets
are thought to be the scars of past planet-planet scattering events
\citep{ada03,cha08,jur08}. We mined a dataset of simulations of
planet-planet scattering to determine what conditions were required to
create such systems.

The simulations we used were presented in several previous studies of
planet-planet scattering and its consequences
\citep{ray08,ray09,ray10,tim13}. We focused on a particular set of
simulations referred to as the {\tt mixed} set in previous papers.
These simulations started from three giant planets with masses between
one Saturn mass and $1000 M_\oplus$ (roughly 3 Jupiter masses), chosen
to follow a $dN/dM \propto M^{-1.1}$ distribution \citep{but06}. The
planets were initially placed on circular orbits with small mutual
inclinations between a few and ten AU, spaced by 4-5 mutual Hill
radii. About 60\% of the simulations became unstable within 100 Myr.

Our sample consists of 448 simulations in which two planets survived
(and energy was adequately conserved). 14 systems (3.1\% of the
sample) contained an outer planet on an eccentric orbit ($e_{outer} >
0.5$) and a significantly less-eccentric inner planet ($e_{inner} <
0.25$). These represent potential analogs to the systems studied in
this paper.

The 14 potential analogs differed from the rest of the sample in two
ways. First, the innermost planet -- both at the start and end of
the simulations -- was systematically more massive than the outer
planet(s). A K-S test showed that the difference between the
potential analogs and the other simulations was statistically
significant, with $p = 3 \times 10^{-3}$. The number of encounters
undergone during the scattering process was higher for the potential
analogs but with a lower statistical significance ($p = 0.03$). In
essence, the analog systems were created by multiple scattering
between planets in the outer parts of their systems and this
scattering was only weakly coupled to the more massive, inner planet.

The mass ratios between the surviving planets are again different
between the control sample and the potential analogs. The potential
analogs have a higher inner-to-outer planetary mass ratio with a
strong statistical significance ($p = 7 \times 10^{-4}$). This is
because a more massive inner planet is less affected by the scattering
between outer planets and thus keeps a smaller eccentricity. However,
the four systems that we study here all have more massive outer
planets.

There was only one analog system with a more massive outer planet. In
that system the planetary instability was system-wide. The inner
planet was initially scattered outward by the middle planet, then back
inward by the outer planet. The planet that was initially the middle
one was then scattered outward, was again repeatedly scattered by the
roughly equal-mass outer planet and eventually ejected from the
system. The outer planet's large eccentricity was a result of its
having been scattered off a similar-sized planet, scattering among
equal-mass planets representing the strongest possible eccentricity
increase \citep{ray10}.

To explain the observed systems we therefore need to invoke strong
scattering events with specific characteristics. The scattering must
have most strongly affected the outermost observed planet and it must
have included a roughly equal-mass planet that was ejected. The
innermost, less massive planet, must have been sufficiently removed
from this scattering to survive on a lower eccentricity orbit (clearly
seen in Figure \ref{comp} by the large orbital separations in these
systems). The most likely origin of such conditions would therefore be
systems in which two or more high-mass planets formed on wider orbits
than an inner, lower-mass planet.

Of course, this assumes a self-unstable system. There is also the
possibility of an outside-in perturbation on the system from a passing
star or wide binary \citep{zak04,mal07,kai13}. In that case, a
significant separation between the inner and outer planets is still
needed to avoid transmitting the perturbation all the way to the inner
parts of the system.


\section{Discussion}
\label{discussion}

\begin{figure*}
  \begin{center}
    \begin{tabular}{cc}
      \includegraphics[angle=270,width=7.5cm]{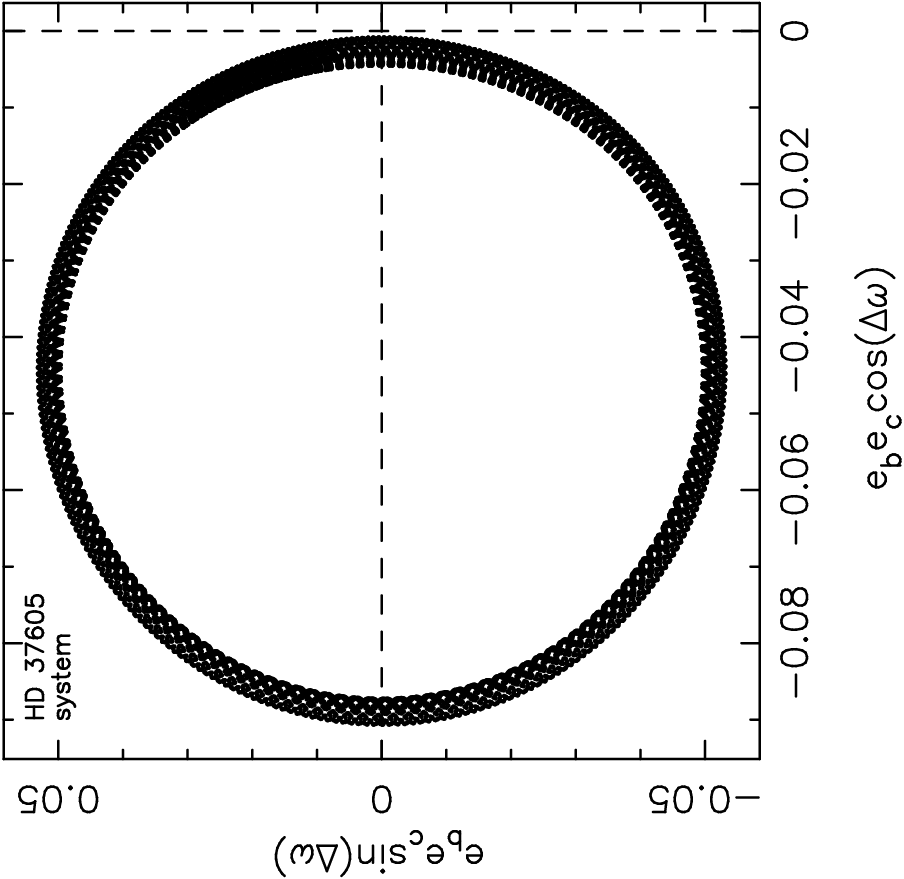} &
      \includegraphics[angle=270,width=7.5cm]{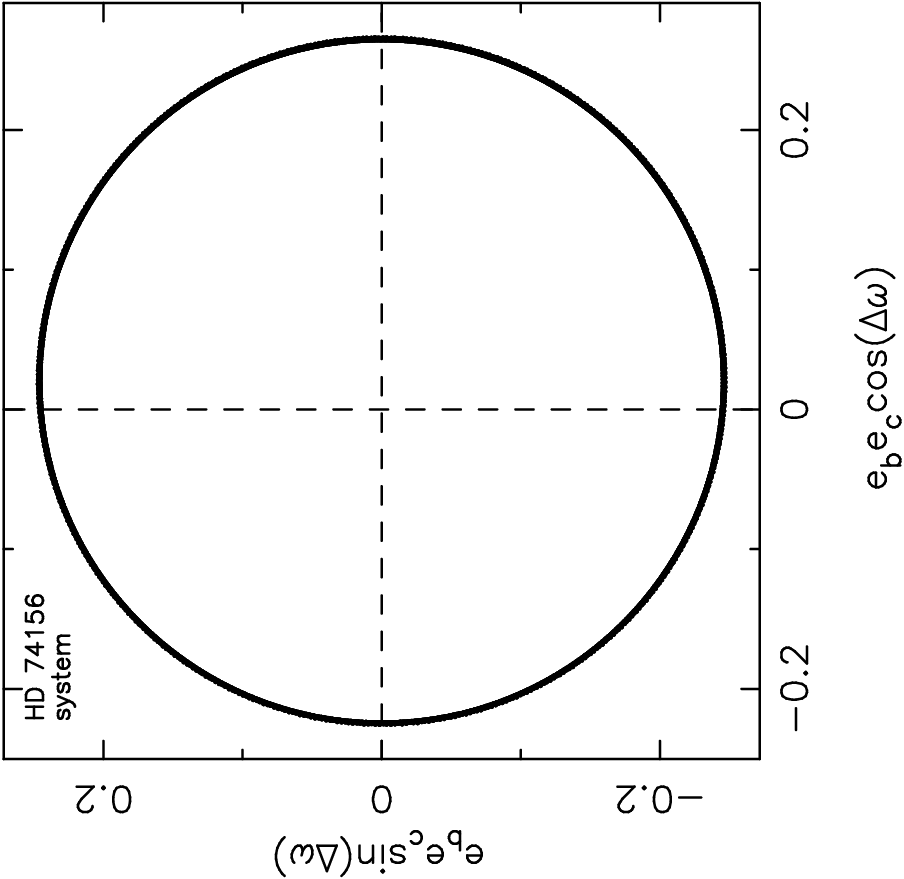} \\
      \includegraphics[angle=270,width=7.5cm]{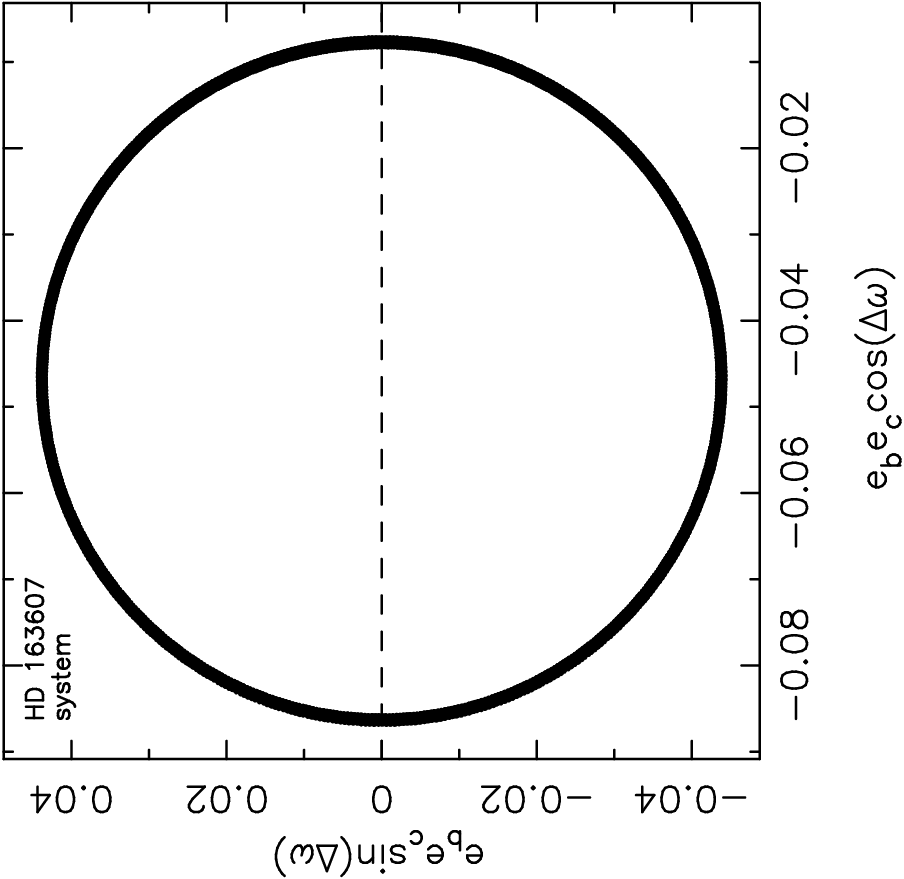} &
      \includegraphics[angle=270,width=7.5cm]{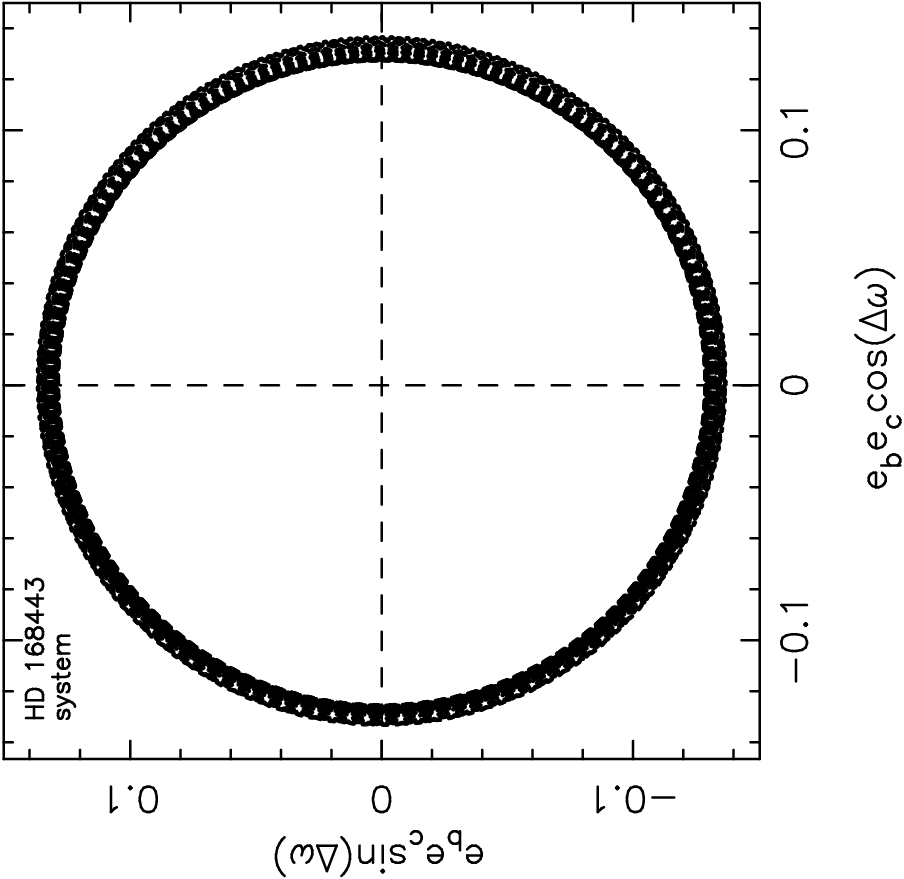}
    \end{tabular}
  \end{center}
  \caption{A polar plot of $e_b e_c$ versus $\Delta \omega$ for each
    of the four systems described in this paper: HD~37605 (top-left),
    HD~74156 (top-right), HD~163607 (bottom-left), HD~168443
    (bottom-right). According to the criteria of \citet{bar06c}, the
    apsidal modes are librating for HD~37605 and HD~163607, and
    circulating for HD~74156 and HD~168443.}
  \label{epsilon}
\end{figure*}

An aspect of interest in the orbital configurations of the systems
studied here is the origin and subsequent evolution of those orbits
(see Section \ref{origin}). \citet{ver09} showed that finding these
kinds of systems at a particular epoch can have a low associated
probability. For example, Figure \ref{hd37605fig} reveals that the c
planet in the HD~37605 system spends the majority of its orbit at a
significantly higher eccentricity than that which is currently
observed. It is also unlikely that the planets would have formed in
such orbits without significant transfers of angular momentum to an
additional component (stellar or planetary) no longer present in the
system. Thus, an important property of these systems with eccentricity
diversity is the angular momentum deficit (AMD), which is the
difference in angular momentum of the system compared with the angular
momentum if the planets were in circular orbits with the same
semi-major axes. This is given by the following equation:
\begin{equation}
  \mathrm{AMD} \equiv \sum\limits_{i=1}^{N} L_p \left( 1 - \sqrt{1 -
    e_i^2} \right)
\end{equation}
where
\begin{equation}
  L_p = \frac{M_{p,i} M_\star}{M_{p,i} + M_\star} \sqrt{G (M_{p,i} +
    M_\star) a_i}
\end{equation}
is the angular momentum of a circular orbit for planet $i$. In cases
where $M_{p,1} \sim M_{p,2}$, such as HD~37605, the inner planet has
much less orbital angular momentum than the outer planet resulting in
larger eccentricity oscillations for a given AMD value.

The system with the smallest eccentricity oscillations is HD~74156. As
noted in Section \ref{hd74156}, this is also the system with the
highest planetary mass ratio and eccentricity of the outer
planet. These facts combined with the relatively slow periastron
precession of the outer planet are consistent with the findings of
\citet{bar04} that the system is unlikely to possess unstable
configurations. Calculations of the AMD for these systems show that
the HD~74156 system has a substantially higher (factor of 2) AMD than
the other three.

The various types of apsidal motion in interacting exoplanetary
systems have been investigated in detail by \citet{bar06a,bar06c}.
\citet{bar06c} distinguish between two basic types of apsidal
behavior, libration and circulation, where the boundary between them
is referred to as a secular separatrix. This distinction may be used
as an additional tool in characterizing the long-term behavior of
Keplerian orbital elements of multi-planet systems. Figure
\ref{epsilon} represents the apsidal trajectory graphically for the
four systems discussed here. The plot uses the eccentricity of the
inner and outer planets ($e_b$ and $e_c$ respectively) and the
difference in periastron arguments ($\Delta \omega$) to create a polar
representation of the apsidal behavior. This shows that the apsidal
modes of HD~74156 and HD~168443 are circulating since the polar
trajectories encompass the origin. This is consistent with the
relatively small amplitude of the c planet eccentricity evolutions
shown in Figures \ref{hd74156fig} and \ref{hd168443fig}. Conversely,
HD~37605 and HD~163607 are shown to be librating systems and are close
to the separatrix between secular libration and circulation. This is
also consistent with Figures \ref{hd37605fig} and \ref{hd163607fig}
since those show that the c planet eccentricities regularly approach
zero.

Finally, it is worth considering as to whether the oscillations
described here in both the eccentricity and periastron precession may
be detectable in reasonable timescales. The high-frequency secondary
oscillations in the eccentricity of the outer planet, such as HD~37605
and HD~168443, are of relatively low amplitude. The orbital
precession of exoplanets has been previously investigated by
\citet{kan12b} who found that the timescales for such precession is
unlikely to allow detection except for the very short-period
planets. By comparison, the perihelion precessions of Mercury and
Earth are 0.159\degr/century and 0.321\degr/century respectively
\citep{cle47}. For the cases presented in Section \ref{dynamics}, the
period of eccentricity oscillation and periastron precession are at
least several centuries in even the most optimistic cases. The main
hindrance however is that the uncertainties associated with those
parameters tend to be comparable to the expected variations.


\section{Conclusions}

Exoplanet discoveries are leading to a diverse range of multi-planet
system configurations. Here we have concentrated on a particular
configuration which consists of two planets in which the inner planet
has an eccentricity larger than 0.5. As shown in Section \ref{system},
these systems bear many similarities, in particular the ratio of the
outer to inner planet semi-major axes which is likely a constraint
imposed by the unique dynamics present in these systems. Through an
examination of the residuals to the radial velocity fits to the data,
we are able to exclude the presence of planets with masses larger than
$\sim 0.5 M_J$ with orbital periods less than 1000 days at the
3$\sigma$ level.

Our N-body integrations of each system show that there is a
synchronous transfer of angular momentum between the planets in most
cases. The case of HD~74156 is interesting due to the high mass ratio,
the relatively small amplitude of eccentricity oscillation and
periastron precession of the outer planet, and the relatively high AMD
for the system. The high-frequency secondary eccentricity oscillations
for the outer planets is indicative of outer planet responses to
frequent dynamical interactions with the much shorter period inner
planet. The effect of introducing a mutual inclination is to increase
the amplitude of secondary oscillations in the inner planet until that
planet is subsequently ejected from the system. The AMDs of these
systems provide indications that these systems have indeed had a
significant perturbation event which produced these relatively rare
orbital configurations. As more of these kinds of systems are
discovered, we can gain further understanding as to their frequency
and dynamical histories.


\section*{Acknowledgements}

The authors would like to thank the anonymous referee, whose comments
greatly improved the quality of the paper. This research has made use
of the Exoplanet Orbit Database and the Exoplanet Data Explorer at
exoplanets.org. This research has also made use of the NASA Exoplanet
Archive, which is operated by the California Institute of Technology,
under contract with the National Aeronautics and Space Administration
under the Exoplanet Exploration Program.



\begin{thebibliography}{}

\bibitem[\protect\citeauthoryear{Adams \& Laughlin}{2003}]{ada03}
  Adams, F.C., Laughlin, G. 2003, Icarus, 163, 290
\bibitem[\protect\citeauthoryear{Akeson et al.}{2013}]{ake13} Akeson,
  R.L., et al. 2013, PASP, 125, 989
\bibitem[\protect\citeauthoryear{Anglada-Escud\'e et
    al.}{2010}]{ang10} Anglada-Escud\'e, G., L\'opez-Morales, M.,
  Chambers, J.E., 2010, ApJ, 709, 168
\bibitem[\protect\citeauthoryear{Barnes \& Quinn}{2004}]{bar04}
  Barnes, R., Quinn, T. 2004, ApJ, 611, 494
\bibitem[\protect\citeauthoryear{Barnes \& Greenberg}{2006a}]{bar06a}
  Barnes, R., Greenberg, R. 2006, ApJ, 638, 478
\bibitem[\protect\citeauthoryear{Barnes \& Greenberg}{2006b}]{bar06b}
  Barnes, R., Greenberg, R. 2006, ApJ, 647, L163
\bibitem[\protect\citeauthoryear{Barnes \& Greenberg}{2006c}]{bar06c}
  Barnes, R., Greenberg, R. 2006, ApJ, 652, L53
\bibitem[\protect\citeauthoryear{Barnes \& Greenberg}{2007}]{bar07}
  Barnes, R., Greenberg, R. 2007, ApJ, 665, L67
\bibitem[\protect\citeauthoryear{Butler et al.}{2006}]{but06} Butler,
  R.P., et al., 2006, ApJ, 646, 505
\bibitem[\protect\citeauthoryear{Chambers et al.}{1996}]{cha96}
  Chambers, J.E., Wetherill, G.W., Boss, A.P. 1996, Icarus, 119, 261
\bibitem[\protect\citeauthoryear{Chambers}{1999}]{cha99} Chambers,
  J.E. 1999, MNRAS, 304, 793
\bibitem[\protect\citeauthoryear{Chatterjee et al.}{2008}]{cha08}
  Chatterjee, S., Ford, E.B., Matsumura, S., Rasio, F.A. 2008, ApJ,
  686, 580
\bibitem[\protect\citeauthoryear{Clemence}{1947}]{cle47} Clemence,
  G.M. 1947, RvMP, 19, 361
\bibitem[\protect\citeauthoryear{Cochran et al.}{1997}]{coc97}
  Cochran, W.D., Hatzes, A.P., Butler, R.P., Marcy, G.W. 1997, ApJ,
  483, 457
\bibitem[\protect\citeauthoryear{Cochran et al.}{2011}]{coc04}
  Cochran, W.D., et al. 2004, ApJ, 611, L133
\bibitem[\protect\citeauthoryear{Ford \& Rasio}{2008}]{for08} Ford,
  E.B., Rasio, F.A., 2008, ApJ, 686, 621
\bibitem[\protect\citeauthoryear{Giguere et al.}{2012}]{gig12}
  Giguere, M.J., et al. 2012, ApJ, 744, 4
\bibitem[\protect\citeauthoryear{Giuppone et al.}{2012}]{giu12}
  Giuppone, C.A., Ben\'itez-Llambay, P., Beaug\'e, C. 2012, MNRAS,
  421, 356
\bibitem[\protect\citeauthoryear{Goldreich \& Soter}{1966}]{gol66}
  Goldreich, P., Soter, S., 1966, Icarus, 5, 375
\bibitem[\protect\citeauthoryear{Goldreich et al.}{2004}]{gol04}
  Goldreich, P., Lithwick, Y., Sari, R. 2004, ApJ, 614, 497
\bibitem[\protect\citeauthoryear{Juri\'c \& Tremaine}{2008}]{jur08}
  Juri\'c, M., Tremaine, S., 2008, ApJ, 686, 603
\bibitem[\protect\citeauthoryear{Kaib et al.}{2013}]{kai13} Kaib,
  N.A., Raymond, S.N., Duncan, M. 2013, Nature, 493, 381
\bibitem[\protect\citeauthoryear{Kane et al.}{2012a}]{kan12a} Kane,
  S.R., Ciardi, D.R., Gelino, D.M., von Braun, K. 2012, MNRAS, 425,
  757
\bibitem[\protect\citeauthoryear{Kane et al.}{2012b}]{kan12b} Kane,
  S.R., Horner, J., von Braun, K. 2012, ApJ, 757, 105
\bibitem[\protect\citeauthoryear{Laughlin \& Chambers}{2001}]{lau01}
  Laughlin, G., Chambers, J.E. 2001, ApJ, 551, L109
\bibitem[\protect\citeauthoryear{Laughlin \& Chambers}{2002}]{lau02}
  Laughlin, G., Chambers, J.E. 2002, AJ, 124, 592
\bibitem[\protect\citeauthoryear{Lin \& Ida}{1997}]{lin97} Lin,
  D.N.C., Ida, S. 1997, ApJ, 477, 781
\bibitem[\protect\citeauthoryear{Malmberg et al.}{2007}]{mal07}
  Malmberg, D., de Angeli, F., Davies, M.B., Church, R.P., Mackey, D.,
  Wilkinson, M.I. 2007, MNRAS, 378, 1207
\bibitem[\protect\citeauthoryear{Malmberg \& Davies}{2008}]{mal08}
  Malmberg, D., Davies, M.B., 2009, MNRAS, 394, L26
\bibitem[\protect\citeauthoryear{Marcy et al.}{1999}]{mar99} Marcy,
  G.W., Butler, R.P., Vogt, S.S., Fischer, D., Liu, M.C. 1999, ApJ,
  520, 239
\bibitem[\protect\citeauthoryear{Marcy et al.}{2001}]{mar01} Marcy,
  G.W., et al. 2001, ApJ, 555, 418
\bibitem[\protect\citeauthoryear{Matsumura et al.}{2008}]{mat08}
  Matsumura, S., Takeda, G., Rasio, F.A., 2008, ApJ, 686, L29
\bibitem[\protect\citeauthoryear{Meschiari et al.}{2011}]{mes11}
  Meschiari, S., Laughlin, G., Vogt, Steven S. Butler, R.P., Rivera,
  E.J., Haghighipour, N., Jalowiczor, P. 2011, ApJ, 727, 117
\bibitem[\protect\citeauthoryear{Naef et al.}{2001}]{nae01} Naef, D.,
  et al. 2001, A\&A, 375, L27
\bibitem[\protect\citeauthoryear{Naef et al.}{2004}]{nae04} Naef, D.,
  Mayor, M., Beuzit, J.L., Perrier, C., Queloz, D., Sivan, J.P., Udry,
  S. 2004, A\&A, 414, 351
\bibitem[\protect\citeauthoryear{Pilyavsky et al.}{2011}]{pil11}
  Pilyavsky, G., et al. 2011, ApJ, 743, 162
\bibitem[\protect\citeauthoryear{Rasio \& Ford}{1996}]{ras96} Rasio,
  F.A., Ford, E.B. 1996, Science, 274, 954
\bibitem[\protect\citeauthoryear{Raymond et al.}{2008}]{ray08}
  Raymond, S.N., Barnes, R., Armitage, P.J., Gorelick, N. 2008, ApJ,
  687, L107
\bibitem[\protect\citeauthoryear{Raymond et al.}{2009}]{ray09}
  Raymond, S.N., Armitage, P.J., Gorelick, N. 2009, ApJ, 699, L88
\bibitem[\protect\citeauthoryear{Raymond et al.}{2010}]{ray10}
  Raymond, S.N., Armitage, P.J., Gorelick, N. 2010, ApJ, 711, 772
\bibitem[\protect\citeauthoryear{Rodigas \& Hinz}{2009}]{rod09}
  Rodigas, T.J., Hinz, P.M. 2009, ApJ, 702, 716
\bibitem[\protect\citeauthoryear{Timpe et al.}{2013}]{tim13} Timpe,
  M., Barnes, R., Kopparapu, R., Raymond, S.N., Greenberg, R.,
  Gorelick, N. 2013, AJ, 146, 63
\bibitem[\protect\citeauthoryear{Veras \& Ford}{2009}]{ver09} Veras,
  D., Ford, E.B. 2009, ApJ, 690, L1
\bibitem[\protect\citeauthoryear{Wang \& Ford}{2011}]{wan11} Wang, J.,
  Ford, E.B., 2011, MNRAS, 418, 1822
\bibitem[\protect\citeauthoryear{Wang et al.}{2012}]{wan12} Wang,
  S.X., et al. 2012, ApJ, 761, 46
\bibitem[\protect\citeauthoryear{Weidenschilling \&
    Marzari}{1996}]{wei96} Weidenschilling, S.J., Marzari, F. 1996,
  Nature, 384, 619
\bibitem[\protect\citeauthoryear{Wisdom \& Holman}{1991}]{wis91}
  Wisdom, J., Holman, M. 1991, AJ, 102, 1528
\bibitem[\protect\citeauthoryear{Wisdom}{2006}]{wis06} Wisdom,
  J. 2006, AJ, 131, 2294
\bibitem[\protect\citeauthoryear{Wittenmyer et al.}{2013}]{wit13}
  Wittenmyer, R.A., et al. 2013, ApJS, 208, 2
\bibitem[\protect\citeauthoryear{Wright \& Howard}{2009}]{wri09}
  Wright, J.T., Howard, A.W. 2009, ApJS, 182, 205
\bibitem[\protect\citeauthoryear{Wright et al.}{2011}]{wri11}
  Wright, J.T., et al. 2011, PASP, 123, 412
\bibitem[\protect\citeauthoryear{Zakamska \& Tremaine}{2004}]{zak04}
  Zakamska, N.L., Tremaine, S. 2004, AJ, 128, 869

\end{thebibliography}
\end{document}